\documentclass[usenatbib,usegraphicx]{mn2e}
\usepackage{natbib}
\usepackage{times}
\newcommand{\vmax}{\mbox{\ensuremath{V_{\mathrm{max}}}}}

\newcommand       \be           {\begin{equation}}
\newcommand       \ee           {\end{equation}}

\def\lsim{\mathrel{\mathpalette\@versim<}}  
\def\gsim{\mathrel{\mathpalette\@versim>}}  

\begin{document}


\title{Luminosity Dependence in the Fundamental Plane Projections of
Elliptical Galaxies}

\author[L.-B. Desroches et al.]{Louis-Benoit
Desroches$^1$\thanks{Email: louis@astro.berkeley.edu}, Eliot Quataert$^1$\thanks{Email: eliot@astro.berkeley.edu}, Chung-Pei Ma$^1$\thanks{Email: cpma@astro.berkeley.edu}, Andrew A. West$^1$\thanks{Email: awest@astro.berkeley.edu} \\
$^1$Department of Astronomy, University of California at
Berkeley, Berkeley, CA 94720-3411 USA}
 
 \date{\it Draft version \today}

\maketitle
 

\begin{abstract}

We analyze the fundamental plane projections of elliptical galaxies
as a function of luminosity, using a sample of $\approx 80,000$
galaxies drawn from Data Release 4 (DR4) of the Sloan Digital Sky
Survey (SDSS).  We separate brightest cluster galaxies (BCGs) from our
main sample and reanalyze their photometry due to a problem with the default pipeline sky subtraction
for BCGs.  The observables we consider are effective radius ($R_e$),
velocity dispersion ($\sigma$), dynamical mass ($M_{\rm dyn} \propto
R_e\sigma^2$), effective density ($\sigma^2/R_{\!e}^2$), and effective
surface brightness ($\mu_e$).  With the exception of the $L$-$M_{\rm
dyn}$ correlation, we find evidence of variations in the slope (i.e. the power-law index) of the
fundamental plane projections with luminosity for our normal elliptical galaxy population.  
In particular, the
radius-luminosity and Faber-Jackson relations are steeper at high
luminosity relative to low luminosity, and the more luminous
ellipticals become progressively less dense and have lower surface
brightnesses than lower luminosity ellipticals. These variations can
be understood as arising from differing formation histories, with more
luminous galaxies having less dissipation. Data from the literature and our reanalysis of
BCGs show that BCGs have radius-luminosity and Faber-Jackson relations
steeper than the brightest non-BCG ellipticals in our sample,
consistent with significant growth of BCGs via dissipationless
mergers.  The variations in slope we find in the Faber-Jackson relation of non-BCGs are
qualitatively similar to that reported in the black hole
mass-velocity dispersion ($M_{\rm BH}$-$\sigma$) correlation.  This
similarity is consistent with a roughly constant value of $M_{\rm
BH}/M_\star$ over a wide range of early type galaxies, where $M_\star$
is the stellar mass.

 \end{abstract}



\section{INTRODUCTION}

Early-type galaxies are observed to populate a tight plane -- the
fundamental plane (FP) -- in the space defined by their effective
radii, velocity dispersions, and surface brightnesses
(\citealt{Dressler87,DD87,BBF92,Burstein97}). Such a correlation is expected
theoretically if early-type galaxies are in virial equilibrium.  The
observed plane is, however, tilted with respect to the simplest virial
theorem expectation, which must be accounted for in galaxy formation models. 
This tilt could be caused by a variation in the dynamical
mass-to-light ratio for elliptical galaxies as a result of a varying dark matter fraction (e.g.,
\citealt{Padmanabhan04,BMQ05}) or stellar population variations (e.g.,
\citealt{Gerhard01}). Another explanation for the FP tilt could be
non-homology in the surface brightness profiles of elliptical galaxies
(e.g., \citealt{GC97,TBB04}).

Projections of the FP contain additional information about the
properties of early type galaxies. These include, e.g., the
Faber-Jackson (FJ) relation \citep{FJ76} between velocity dispersion and
luminosity and the radius-luminosity relation.  Although the FP itself
is quite tight over a wide range of early type galaxies (e.g.,
\citealt{Bernardi03c}), there are a number of indications of
variations in the FP projections themselves with galaxy luminosity (which we refer to as ``variations in slope'' or ``curvature'' in the FP projections).
For example, the slope of the Faber-Jackson relation decreases in
lower luminosity ellipticals \citep{Tonry81,Davies83,MG05}.  In addition,
\citet{OH91} find that while brightest cluster galaxies (BCGs) lie on
the same FP as lower mass ellipticals, their central velocity
dispersions are roughly constant with increasing luminosity and their
effective radii increase steeply with luminosity, significantly more
so than for normal ellipticals. \citet{Lauer06a} and \citet{Bernardi06}
have confirmed these differences in the FP projections of BCGs with
larger data sets. These trends can be understood if dissipationless (gas-poor) mergers become increasingly important for luminous ellipticals \citep{BMQ06}.

In addition to global correlations among the galaxy properties
themselves, the masses of central black holes in early-type galaxies
correlate well with the velocity dispersions
\citep{FM00,Gebhardt00,Ferrarese01,Tremaine02}, bulge concentrations \citep{Graham01}, and stellar masses
\citep{Magorrian98,MH03,HR04} of their host galaxies. \citet{Wyithe06} and \citet{GH06}
have argued that the $M_{\rm BH}-\sigma$ relation steepens at high
black hole masses and flattens at low masses, respectively. This is similar to the
luminosity dependence in the FJ relation noted above. If supported by
larger data sets, the similar luminosity dependence of the $M_{\rm
BH}-\sigma$ and FJ relations could have important implications for
understanding the relationship between the formation of galaxies and
their central massive black holes. 

The purpose of the present paper is to quantify the variation in the
FP projections with luminosity using the exquisite statistics made
possible by the Sloan Digital Sky Survey (SDSS) Data Release 4 (DR4;
\citealt{sdss_dr4}). These variations have, until now, been largely overlooked by previous FP studies, because simple power-law assumptions were often made when fitting the data. This can be seen, for instance, in the sample of $\sim9000$ galaxies from \citet{Bernardi03b}, where hints of the variations we discuss are quite noticeable in the plots. In the next section we describe the data used in
our analysis (\S \ref{sec:data}). We also describe a problem in the
standard pipeline SDSS photometry of BCGs that leads us to reanalyze the
BCGs separately from the bulk of the ``normal'' elliptical galaxy population.
Having defined our sample, in \S \ref{sec:FP} we examine the observed
FP projections and quantify curvature in these projections as a
function of luminosity. We also use published
photometry of nearby (non-SDSS) BCGs to discuss the relationship
between our results and the properties of BCGs.  In \S \ref{sec:disc} we discuss our results
and their connection to galaxy-BH correlations.  
Throughout this paper, we use a standard cosmology with
$(\Omega_M,\Omega_\Lambda,h)=(0.3,0.7,0.7)$.

\section{DATA}
\label{sec:data}

The data used in this paper were obtained from the NYU Value Added
Galaxy Catalog (VAGC) \citep{Blanton05}, which uses the SDSS Data Release 4 \citep{sdss_dr4}. This
includes both the imaging and spectroscopic catalogs, as well as the
\citet{Sersic63,Sersic68} model fitting parameters \citep{Blanton03_apj594} in
5 bands ({\it ugriz}). We also use the VAGC $K$-corrections (version
\verb!4_1_4!)  \citep{Blanton03_aj125}. SDSS magnitudes are calibrated
to the AB system \citep{OG83}, in which a magnitude 0 object has the
same counts as a $F_\nu=3631$ Jy source (this zero point has been
confirmed in the {\it r} band). We use SDSS magnitudes $K$ corrected to redshift $z=0$. We restrict the sample to $15 < $ {\it
r} $ < 17.5$, where {\it r} is the \citet{Petrosian76} {\it r}-band
extinction-corrected apparent magnitude. These bright and faint target
limits define a complete sample (target selection is done only in {\it
r}). The faint limit is {\it r} $=17.77$ in some parts of the sky, but
to be safe, we adopt 17.5. We also only consider galaxies with
redshift $0.01 < z < 0.3$. The low redshift cutoff excludes galaxies
with large peculiar velocities. We primarily present results in the
{\it r} band, but note that the {\it g} and {\it i} bands show very
similar behavior. The {\it u} and {\it z} bands suffer from filter and
chip problems, as well as low signal levels, and so we do not consider
them.

In order to isolate elliptical galaxies, we use the following three
main criteria. First, the S\'ersic index in {\it gri} must satisfy $n
> 2.5$, in order to isolate \citet{deV48} profiles ($n=4$) over
exponential profiles ($n=1$). Second, we restrict the concentration
index $c = R_{90}/R_{50} > 2.86$ in {\it r}, where $R_{50}$ and
$R_{90}$ are the Petrosian 50\% and 90\% light radii, respectively
(for reasons explained shortly, we also require the concentration
index $c < 3.8$ in {\it gri}).  The criterion $c > 2.86$ separates
early and late-type galaxies with a completeness of $\sim$82\%
\citep{Nakamura03,Shimasaku01}. The concentration is not corrected for
seeing effects, and thus a compact galaxy observed under bad seeing
conditions will have an underestimated concentration
\citep{Blanton03_apj594}. The sample of \citet{Nakamura03} has eyeball
classifications, however, and the cutoff $c=2.86$ was chosen to
separate early and late types with minimal cross-contamination. For
comparison, the S\'ersic index above does take into account the
observed point spread function (PSF).  In addition to the above cuts
on the surface brightness profile, we also apply the colour criterion
$^{0.1}(${\it g-r}$)>0.7$, where $^{0.1}(${\it g-r}$)$ is the {\it
g-r} colour $K$-corrected to redshift 0.1. \citet{Blanton03_apj594}
found that this criterion isolates early-type galaxies from spirals.
We note that the three criteria above are not completely
independent. The most restrictive is the concentration cut, followed
by the S\'ersic cut and the colour cut. Their combination allows us to
remove marginal outliers, especially S0 galaxies, and thus focus on
the normal elliptical population.

In addition to the primary selection criteria above, several
additional criteria are used to remove bad points, notably using
pipeline warning flags, requiring a median signal-to-noise (S/N) $>10$
in the spectrum, eliminating galaxies with very high or low
(unphysical) values for $\sigma$, requiring the de Vaucouleurs
likelihood of the SDSS model fit to be greater than the exponential
likelihood, and ensuring that all quantities have well-defined
measurement errors. These quality restrictions remove roughly 7\% of
galaxies identified by our elliptical criteria. For reasons explained
in \S \ref{sec:data_bcg}, we also separate out all BCGs identified in the
C4 catalog of \citet{Miller05} from our main sample.  The final sample
(hereafter known as the main normal elliptical galaxy sample) consists of 79,482 galaxies.

The VAGC provides three different magnitude definitions for galaxies:
a) a Petrosian magnitude, b) a magnitude derived from a
S\'ersic fit, and c) a magnitude derived from a de Vaucouleurs
fit. We elect to use Petrosian magnitudes, converted into total
``S\'ersic''-like magnitudes using the concentration index $c$ and the
relations of \citet{Graham05} for SDSS data. In particular, \be
M_{tot} \approx M_P-P_1\exp (c^{P_2}), \ee where $M_{tot}$ is the
total ``S\'ersic''-like magnitude, $M_P$ the Petrosian magnitude, and
$c$ the concentration. $P_1=4.2 \times 10^{-4}$ and $P_2=1.514$ for
SDSS data.  These transformations effectively recover the magnitudes
that would be obtained from an ideal S\'ersic fit using only Petrosian
quantities. The only assumption needed is that a galaxy's profile is
well modeled by a S\'ersic function, although a S\'ersic fit is not
actually used. The \citet{Graham05} relations also yield an effective
radius $R_e$ (i.e., the half-light radius), using \be R_e \approx
\frac{R_{50}}{1-P_3 c^{P_4}}, \ee where $R_{50}$ is the Petrosian
half-light radius. $P_3=6.0 \times 10^{-6}$ and $P_4=8.92$ for SDSS
data.  The above transformation equations break down for $c > 3.8$, so
we exclude such galaxies. This requirement removes less than 0.5\% of
galaxies, independent of luminosity, and does not effect our results.

The conversion to an effective S\'ersic magnitude using Petrosian
quantities is desirable because (1) de Vaucouleurs magnitudes presume
that all ellipticals are well-modeled as a $R^{1/4}$ profile.  In
fact, however, the S\'ersic index appears to increase mildly with
increasing luminosity (see, e.g.,
\citealt{Caon93,Graham96,Ferrarese06}; we also see the same trend in
our data; see \S \ref{sec:proj}), and (2) Petrosian and S\'ersic
quantities themselves suffer from mild systematic errors.  The
fraction of total flux contained within the Petrosian aperture is a
function of the profile shape (see \citealt{Graham05} for an in-depth
discussion).  For example, the Petrosian aperture contains $\sim98$\%
of the total flux for an exponential profile, but only $\sim80$\% for
a de Vaucouleurs profile.  This introduces a luminosity-dependent
error into the total magnitude definition, given the mild increase in
S\'ersic index with increasing luminosity.  S\'ersic magnitudes
themselves do account for profile shape, and represent a total
integrated luminosity. There are, however, systematic problems with
VAGC S\'ersic fits. As both \citet{Blanton05} and \citet{Graham05}
explain, the S\'ersic fluxes obtained from these fits are
systematically underestimated, and become worse for increasing
S\'ersic index $n$. At $n=5$, the error is $\sim$10\%. S\'ersic fits
are also more sensitive to the radial range over which the fit is
carried out.

For the reasons above, we elect to present results using the total
S\'ersic-like magnitudes derived from Petrosian quantities.  We note,
however, that the entire analysis in this paper was repeated using the
quantities obtained from the S\'ersic and de Vaucouleurs fits. We find
results similar to those presented here, in particular, the same
trends in the FP projections (see \S \ref{sec:proj}). 

The S\'ersic-like Petrosian quantities we use do not take into account possible isophote ellipticity. The majority of galaxies in our sample, however, have axis ratios greater than $\sim 0.6$, and there is no strong dependence on absolute magnitude. As a check, we applied a correction factor from \citet{West05} to our adopted radii and repeated our analysis below. Our measured FP projection slopes changed by at most $\sim 5\%$, with all the same trends being present. Ellipticity therefore does not have an important effect on our results.

In order to correct for passive evolution of early-type galaxies over
the redshift range probed by SDSS, we use the prescription of
\citet{Lin99} to correct all magnitudes; \be M_r(z=0)=M_r(z)+Qz, \ee
where $M_r(z=0)$ is the {\it r}-band absolute magnitude corrected for
evolution, $M_r(z)$ the observed absolute magnitude, $z$ the redshift,
and $Q$ the evolution parameter in magnitudes per redshift. We use the
value of $Q=0.85$ found by \citet{Bernardi03b} for elliptical galaxies
in the {\it r} band. Our net $K$+evolution correction is very similar
to that used by \citet{Wake06} for luminous red galaxies in SDSS in
the {\it r} band, where $K$ is the $K$ correction at redshift $z=0$.

All galaxies in the main sample have a Petrosian $R_{50} > 1\arcsec$,
larger than the typical seeing for the SDSS survey, which is a full
width at half maximum of 1\farcs4 in the {\it r} band.  Thus none of
the measured radii are severely compromised by the PSF. 
As a quantitative check, we repeated our analysis of the FP
projections imposing an additional cut that all galaxies have
$R_{50} > 1\farcs5$, and found quantitatively similar results.

We correct all velocity dispersions for aperture effects following
\citet{JFK95} and \citet{Wegner99}. This corrects all measured
velocity dispersions to a standard circular aperture with a radius
equal to one-eighth the effective radius of the galaxy: \be
\sigma_{\rm{corr}}=\sigma_{\rm{meas}}\left(
\frac{R_{\rm{fiber}}}{R_e/8}\right)^{0.04}, \ee where $\sigma_{\rm{meas}}$ and
$\sigma_{\rm{corr}}$ are the measured and corrected velocity dispersions
respectively, $R_{\rm{fiber}}$ the fiber radius from the SDSS (1\farcs5),
and $R_e$ the effective radius of the galaxy in arcseconds. We note
that this correction assumes that all early-type galaxies have similar
velocity dispersion profiles, regardless of $R_e$.
\citet{Bernardi03a,Bernardi03b} provide a good discussion on this
topic, but given that the correction depends very weakly on $R_e$,
these differences do not have any significant impact on our results.

Finally, we define the effective surface brightness of a galaxy as
follows, $K$-corrected and corrected for cosmological surface
brightness dimming, \be \mu_e=m+2.5\log_{10}(2\pi
R_{\!e}^2)-10\log_{10}(1+z)-K, \ee where $m$ is the extinction- and
evolution-corrected apparent magnitude, $R_e$ the effective radius (in
arcseconds), $z$ the redshift and $K$ the $K$-correction at redshift
$z=0$.

\subsection{Brightest Cluster Galaxies}
\label{sec:data_bcg}

As both \citet{Lauer06a} and \citet{Bernardi06} discuss, there is a
problem with the standard SDSS photometry of luminous BCGs (see
also http://www.sdss.org/dr4/help/known.html).  The problem is that
the default sky subtraction removes the outer low surface brightness
flux from the galaxy, resulting in an underestimate of the luminosity
and effective radii.  We have independently come to the same
conclusion.  This problem is severe for BCGs because most of the
luminosity is in low surface brightness emission.  We were motivated
to look into this problem by the lack of curvature in the initial
\citet{Bernardi03b} SDSS FP projections at high luminosity, relative
to the local BCG samples of \citet{OH91} and Lauer (private
communication \& \citealt{Lauer06a}).

We identify BCGs in SDSS using the C4 catalog of \citet{Miller05},
which is based on DR2 data, complete to $z=0.15$. Clusters are defined in C4 on the basis of position, redshift and colour, and the BCG is simply the brightest system in the cluster. In order to identify
the C4 BCGs in the VAGC catalog, we use their declination and right
ascension to find unique matches to within 1\arcsec \ in the DR4
catalog. We were able to successfully match 744 out of 748 BCGs (some
of which are duplicate objects), although 194 were too bright to be
included in the spectroscopic survey (i.e., brighter than the fiber
saturation limit on SDSS). Of the remaining 550 BCGs, only 346 unique
objects pass all our criteria in establishing an early-type galaxy
sample, the most restrictive being the concentration cut.  We note
that \citet{Bernardi06} also find significant contamination from non
early-type galaxies in the \citet{Miller05} BCG catalog. The majority of this contamination is likely due to a fiber collision restriction in SDSS data, such that the true BCG in a C4 cluster is missed in the spectroscopic catalog \citep{Miller05, vonderLinden06}.

To assess the accuracy of the standard pipeline photometry, we have
reanalyzed images of a subset of high luminosity normal galaxies and
BCGs. Sky subtraction was performed on each individual SDSS field
containing a sample galaxy.  All objects in the field were masked out
and a tilted plane was fit to the remaining sky pixels.  For more
details see \citet{West06}.  In the case where a sample galaxy crossed
multiple fields, the fields were sky subtracted and then mosaicked. We
then performed a 2-dimensional Levenberg-Marquardt least-squares de Vaucouleurs
fit, masking out nearby stars, galaxies, and the center to safely
avoid the PSF core.

Unlike the bulk of the normal elliptical galaxy population, we adopt a de Vaucouleurs fit instead of a S\'ersic fit for BCGs because of the subtle nature of BCG photometry. S\'ersic fits, with an extra degree of freedom, are sensitive to the range of the fit and tend to result in very large luminosities and radii for many BCGs with deep photometry (see \S \ref{sec:disc_bcg}). Much of this flux, however, is probably better attributed to intracluster light rather than host galaxy light, though a proper decomposition into the two components is difficult and will likely require improved theoretical models. In the absence of a more appropriate choice, we adopt a de Vaucouleurs fit for the BCGs.

\begin{figure}
\includegraphics{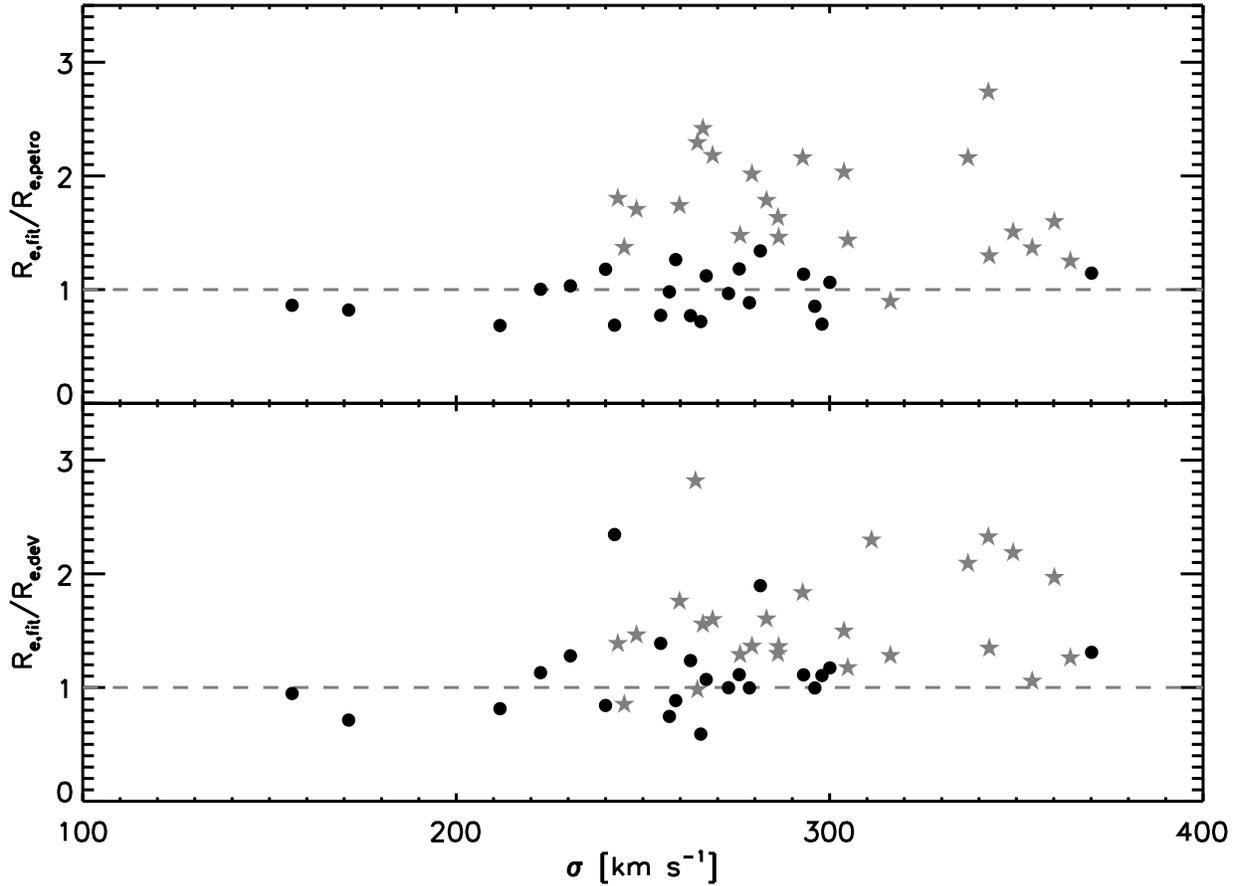}
\caption{A comparison of our fitted $R_e$ to the catalog $R_e$ as a
function of $\sigma$. $R_{e,{\rm fit}}$ is the effective radius from a
de Vaucouleurs fit to a reanalysis of SDSS photometry using
\citeauthor{West06}'s \citeyearpar{West06} sky subtraction. $R_{e,{\rm
petro}}$ is the effective radius derived from catalog S\'ersic-like
Petrosian measurements, as described in \S \ref{sec:data}. $R_{e,{\rm
deV}}$ is the effective radius derived from catalog de Vaucouleurs
fits. Black points are a random sub-sample of high-$L$ normal
ellipticals. Grey stars are a random sub-sample of high-$L$ BCGs. The
BCG effective radii (and luminosities) are consistently
under-estimated with standard SDSS photometry, while most normal
elliptical galaxies appear fine. Two BCGs have $R_{e,{\rm
fit}}/R_{e,{\rm petro}}=4.0$; these are omitted for clarity.}
\label{fig:RevsRe}
\end{figure}

Figure~\ref{fig:RevsRe} shows a plot of the ratio of our fitted de
Vaucouleurs effective radii to the catalog radii as a function of
velocity dispersion $\sigma$ for a sample of normal elliptical
galaxies (black points) and BCGs (grey stars). We include a
comparison to radii obtained from both the S\'ersic-like Petrosian
measurements used in this paper and catalog de Vaucouleurs fits. For
most of the normal galaxies the agreement is quite good, even for very
massive and luminous galaxies with $\sigma \approx 300$ km
s$^{-1}$. This indicates that, for most normal galaxies, the default
catalog photometry is accurate. For the BCGs (gray stars),
however, the catalog effective radii are smaller than our fitted
values by factors of $\approx$1.5--3.  Similarly the total
luminosities of these galaxies are underestimated in the VAGC by
factors of $\approx$1.2--2.5. For these reasons, we exclude BCGs
identified by \citet{Miller05} from our main sample, and reanalyze them separately.

Despite the systematic underestimates in the catalog $R_e$ and
luminosities for SDSS BCGs, BCGs represent only a small fraction of
all elliptical galaxies and so their effect on our main sample as a whole
is small. Indeed, only 346 BCGs identified by \citet{Miller05} in DR2
are excluded from our main sample.  As an independent check on this
number, we note that using the number density of BCGs of
$1.5\times10^{-5} h^3$ Mpc$^{-3}$ at $z \le 0.05$ from \citet{PL95},
we would expect $\approx 400$ BCGs in DR2, in reasonable agreement
with the above numbers.  For comparison, DR2 contains a total of
roughly 10000, 3100, and 400 elliptical galaxies that meet our
criteria brighter than $M_r=-22.5$, $M_r=-23$, and $M_r=-23.5$,
respectively.  Thus even if all BCGs were brighter than $M_r = -23$,
which they are not, BCGs would only become statistically significant
(by number) for galaxies brighter than $M_r \approx -23$. The
precise luminosity at which BCGs become statistically significant
depends, however, on how the luminosity of a BCG is defined, i.e., how
much of the extended low surface brightness emission is detected and
how much is attributed to the galaxy rather than intra-cluster light
(see \S \ref{sec:disc_bcg}). 

To verify the small effects that BCGs have on our results for the normal elliptical galaxy sample, we
performed our analysis of the FP projections on four data sets: a) DR4
with DR2 BCGs excluded (our standard main sample); b) DR4 with no DR2
BCGs excluded; c) DR2 with DR2 BCGs excluded; and d) DR2 with no DR2
BCGs excluded.  In all cases, our $L(\sigma)$ distributions and fits
described below were virtually unchanged. The $L(R_{\!e})$
distributions and fits showed only small differences at the luminous
end. These differences were of order 3\%, roughly equal to our adopted
$1\sigma$ errors at the luminous end. Because the identification of
BCGs by \citet{Miller05} is likely incomplete, we cannot rule out that
contamination by BCGs is present, particularly because we are using
DR4, and \citet{Miller05}'s identification of BCGs is only available
for DR2. Since DR4 contains roughly twice as many galaxies as DR2, we estimate that the number of remaining BCGs in our sample is similar to the number we have excluded ($\sim 350$).
The above comparisons between samples with and without DR2
BCGs show, however, that although the photometry of some remaining
BCGs in our sample may be incorrect, they are not causing significant
errors in our analysis of the FP projections of the non-BCG elliptical
population.  

\section{FUNDAMENTAL PLANE PROJECTIONS}
\label{sec:FP}
\subsection{Magnitude-Limited Sample Corrections}

Our sample is magnitude limited, as defined in Section \ref{sec:data},
at both the bright and faint ends. In order to analyze the
correlations between observables for the elliptical population as a
whole, we must correct for this selection effect. To accomplish this,
we use the \vmax \ method, which weighs each point in the sample by
$1/\vmax$, where \vmax \ is the maximum spatial volume one can observe
a given galaxy in (adjusted for our redshift limits). We ignore
$K$-correction and luminosity evolution in determining \vmax . They
have nearly equal and opposite effects on the observed brightness of a
galaxy (roughly one magnitude per unit redshift), and thus do not
affect our results significantly
\citep{Shen03,Blanton03_aj125,Blanton03_apj592}.  We choose the
nonparametric \vmax \ method over a maximum likelihood analysis to
avoid any assumptions about parametric forms for the intrinsic
distributions, even though we find them to be reasonably Gaussian.  We
note that the \vmax \ method is subject to inaccuracies at the faint
end of the galaxy distribution, where an inhomogeneous and
non-representative sample is given a very high weight. This well-known
problem is addressed below.

As discussed in \citet{Sheth03} using a $1/\vmax$ correction
represents the joint distribution of an observable $X$ and luminosity
$L$. The fit to this joint distribution is known as a bisector fit. In
order to consider the distribution of the observable $X$ at fixed $L$
(i.e., $<\!X\!\mid\!L\!> \propto\!L$), we must weigh each galaxy by
$1/[\phi(L)\vmax]$, where $\phi(L)$ is the luminosity function. The
fit to this distribution is the inverse fit (i.e., $X$ as a function
of $L$).  The inverse fit is more appropriate for these studies
because the observable $X$ is usually far more uncertain than the
luminosity $L$. Treating $L$ as an independent variable and assuming
the scatter is largely due to $X$ is thus the best course of
action. 

One difficulty with the $1/[\phi(L)\vmax]$ correction is that a few
galaxies are given high weight at both the faint and bright ends. We
circumvent this issue by trimming our sample at the bright and faint
ends.  While this somewhat diminishes our results on the curvature in
the FP projections, which are more pronounced at the extreme
luminosity ends, it is necessary to prevent spurious fits dominated by
small number statistics. Our criterion used to trim the sample is that
we consider only galaxies with a weight $W_i=1/[\phi(L_i)\vmax_{,i}]$
no larger than 10$^{1.1}$ times the mean, i.e.,
\be
\frac{W_i}{\langle W_i \rangle} \leq 10^{1.1}.
\label{eq:weight}
\ee
This effectively cuts out galaxies fainter than $M\sim-19.9$ and
brighter than $M\sim-23.7$ in {\it r} band, removing a total of
$\approx 1.7 \%$ of galaxies. This cut is based on trial and error.
Factor of few variations about the choice in equation (\ref{eq:weight})
does not change any of our conclusions, while for a substantially more
permissive cut the fits become overly dominated by a few galaxies,
resulting in very unphysical fits.

\subsection{Observed FP Projections}
\label{sec:proj}

\begin{figure}
\includegraphics{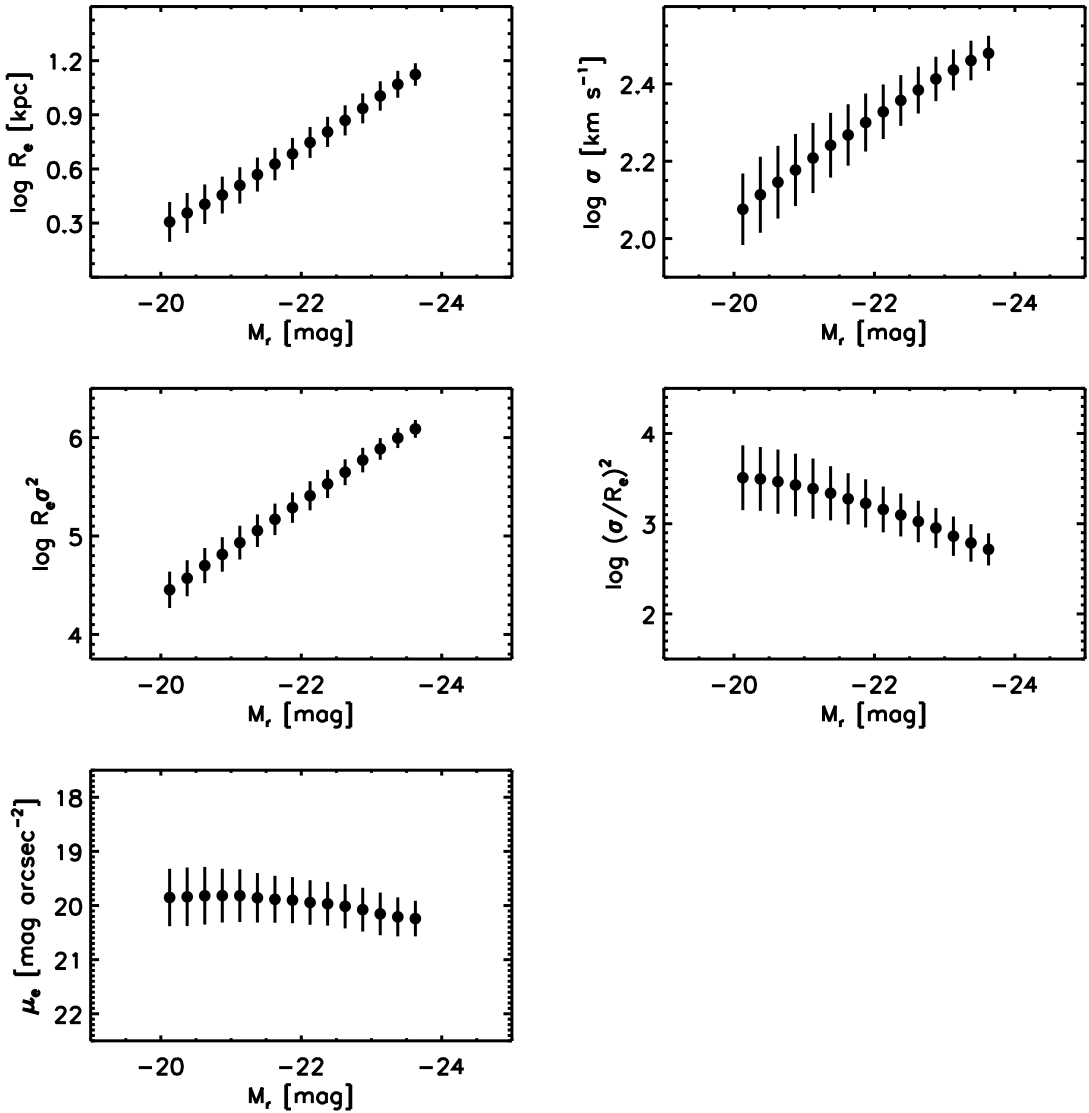}
\caption{Distributions of effective radius $R_e$, velocity dispersion
$\sigma$, dynamical mass $R_e\sigma^2$, effective density
$(\sigma/R_e)^2$, and effective surface brightness $\mu_e$ in 0.25 mag
wide bins, for a sample of 79,482 non-BCG early-type galaxies drawn from SDSS
DR4 that pass concentration, S\'ersic, and colour cuts. Points denote
the peak of a gaussian fit to the distribution in a given bin, while
the error bars correspond to the $1\sigma$ width of the distribution.}
\label{fig:dists}
\end{figure}

Figure~\ref{fig:dists} shows the distribution of observables at fixed
luminosity. The observables we consider are effective radius ($R_e$),
velocity dispersion ($\sigma$), dynamical mass ($M_{\rm dyn} \propto
R_e\sigma^2$), effective density ($\sigma^2/R_{\!e}^2$), and effective
surface brightness ($\mu_e$). We concentrate our discussion on the
first two projections, but include all five for completeness.  In
Figure \ref{fig:dists}, we separate the data points into small 0.25
magnitude-wide bins and show the mean and $1\sigma$ width of the
distribution in each magnitude bin; the distributions are well fit by
Gaussians.

\begin{figure}
\includegraphics{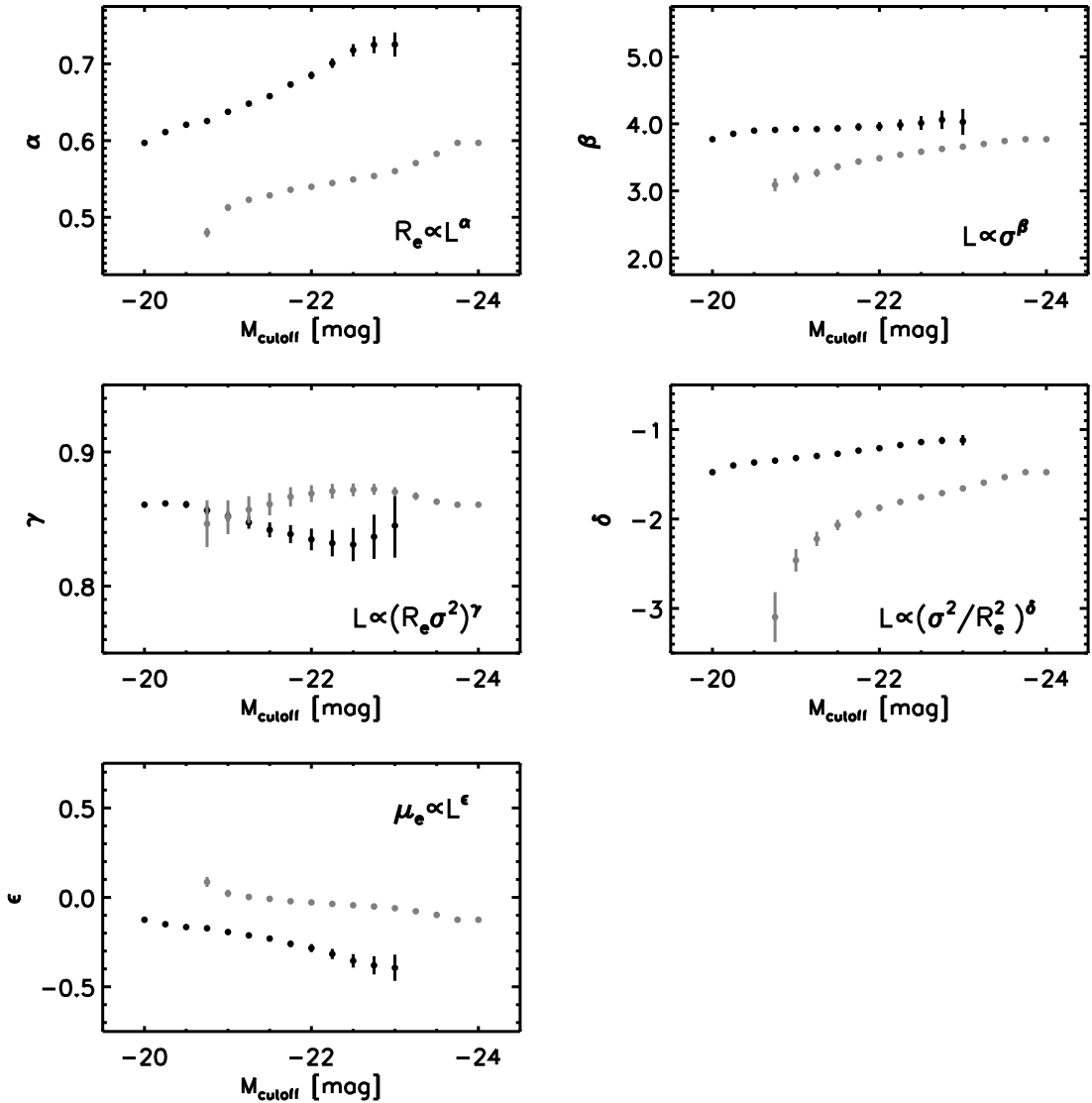}
\caption{Slope of power-law fits for five fundamental plane
projections of our non-BCG elliptical galaxy sample as a function of r-band magnitude cutoff. Black points
represent samples with a lower magnitude cutoff $M_{\rm cutoff}$,
where the sample is increasingly restricted to high luminosities. Grey
points represent samples with an upper magnitude cutoff $M_{\rm
cutoff}$, with the sample increasingly restricted to low
luminosities. These fits show that, with the exception of $L$
vs. dynamical mass ($R_e \sigma^2$), the slopes of the FP projections
depend on the luminosity of the elliptical galaxy sample.}
\label{fig:allslopes}
\end{figure}

Even by eye, Figure \ref{fig:dists} shows pronounced curvature in the
FP projections, with the exception of $R_e\sigma^2$.  We perform two
tests to quantify this curvature.  First, we perform a series of
linear least-squares fits of observable $X$ as a function of magnitude;
we repeat these fits considering different subsamples with varying
bright and faint end magnitude cutoffs.  This quantifies how the slope
of the FP projection depends on the luminosity range considered.
Figure \ref{fig:allslopes} shows the resulting variation in the slope
of the FP projections, considering limits to the sample at the faint
(black points) and bright (grey points) ends.  The left-most black
points and right-most grey points in any of the plots give the slope
of the projection over the entire sample of elliptical galaxies.
The remaining points describe the slope of the FP projections over
increasingly smaller sub-samples. The power-law slopes are defined as follows:
$R_e \propto L^\alpha$, $L \propto \sigma^\beta$, $L \propto (R_e \sigma^2)^\gamma$, $L \propto (\sigma^2/R_{\!e}^2)^\delta$, and $\mu_e \propto L^\epsilon$. The slope error is a statistical
error using the measurement errors on $\sigma$ and $R_e$.

\begin{figure}
\includegraphics{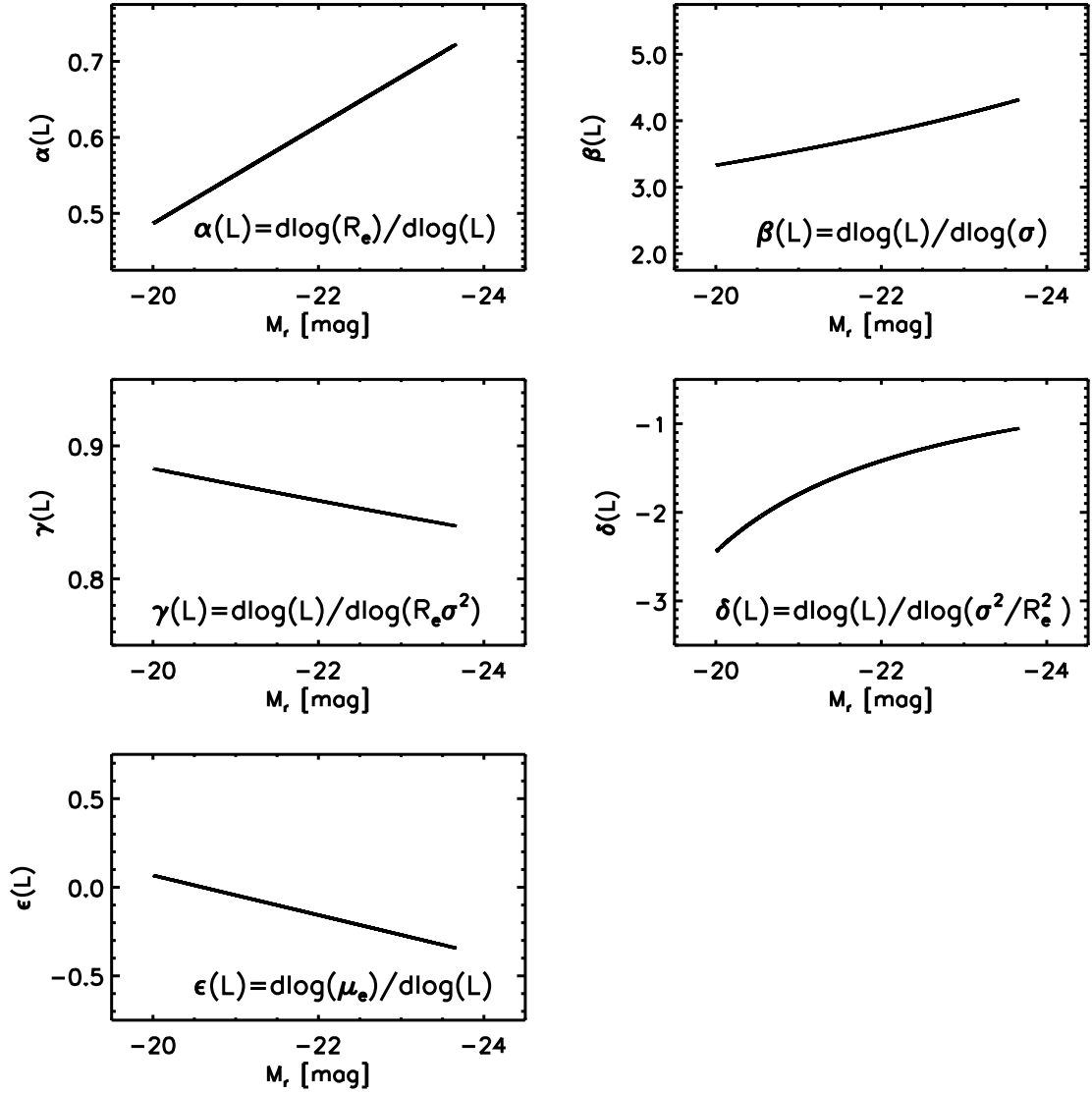}
\caption{Local slope of the fundamental plane projections of our non-BCG elliptical galaxy sample as a
function of $r$-band magnitude. The data in Figure \ref{fig:dists} are
fit to a quadratic function of the form $X = a + b(\log(L)) + c(\log(L))^2$ and the
local slope is defined by the derivative of the fit with respect to
magnitude. Fit parameters are given in Table~\ref{tab:quad}. These results for the variations in the fundamental plane
projections with luminosity are very similar to those of Figure
\ref{fig:allslopes}.  The statistical errors on the slopes are similar
to those in Figure~\ref{fig:allslopes} for the largest subsamples and
are omitted for clarity.}
\label{fig:quad}
\end{figure}

As an alternative way of quantifying curvature in the FP projections,
we consider a quadratic fit to the entire sample, of the form \be X =a
+ b(\log(L)) + c(\log(L))^2,
\label{eq:quad} 
\ee where $L$ is the luminosity and $X$ is the observable.  The
parameters for these fits are presented in Table~\ref{tab:quad}. In
four of the five projections, the quadratic term is statistically
non-zero. The observable $R_e\sigma^2$ is the only one without a clear
detection of a quadratic term. Note that we are not ascribing any
particular physical significance to this second order term, but merely
using it to quantify curvature. By using the derivative of the
quadratic fit with respect to $\log(L)$, we define the local slope of the
$L$-$X$ relation for all observables $X$. This is presented in
Figure~\ref{fig:quad}. The variations in these $L$-$X$ slopes agree
well with the trends seen in Figure~\ref{fig:allslopes}. The
statistical errors on these slopes are similar to those in
Figure~\ref{fig:allslopes} for the largest subsamples and are omitted
for clarity.  

To test the robustness of our local $L$-$X$ slopes, we also performed
3rd order polynomial fits on all our relations, of the form $X = a +
b(\log(L)) + c(\log(L))^2 + d(\log(L))^3$. The best fit local slopes
of the $L$-$X$ relations were nearly identical to our quadratic fits
over the luminosity range of our sample, providing some confidence in
the robustness of the results in Figure~\ref{fig:quad}.  In the case
of the FJ relation, however, a quadratic expansion with respect to
$\log(L)$ might not be the most appropriate description of the data,
given the observed sign of the curvature in Figures \ref{fig:dists} \&
\ref{fig:allslopes}. In equation (\ref{eq:quad}), $\sigma \rightarrow
0$ as $L \rightarrow \infty$ (because $c < 0$ in
Table~\ref{tab:quad}), which is clearly unphysical. An a priori more
reasonable expansion of the FJ relation would be $\log(L)=a+bX+cX^2$,
where $X=\log (\sigma)$ (see also \citealt{Wyithe06}, who carried out
a similar fit for $M_{\rm BH}$-$\sigma$). This now has the correct
asymptotic behavior at large $\sigma$. The resulting best fit local
$L$-$\sigma$ slopes are again almost identical to the results shown in
Figure \ref{fig:quad} over the range in $L$ we are interested in.

It is clear from Figures~\ref{fig:allslopes} and~\ref{fig:quad} that
substantial curvature exists in the FP projections across the entire
sample of normal (non-BCG) elliptical galaxies, with the exception of the dynamical
mass vs. luminosity relation which is well-fit by a single power-law
with $L \propto M_{\rm dyn}^\gamma$, with $\gamma\approx0.86$, which
implies $M_{\rm dyn}/L \propto L^{0.16}$.  This dynamical
mass-to-light ratio agrees well with the DR2 results of
\citet{Padmanabhan04}.  \citet{Padmanabhan04} also show that the
stellar mass-to-light ratio $M_\star /L$ of early type galaxies is
constant with luminosity, using stellar masses determined by
\citet{Kauffmann03}.  Thus $\gamma \ne 1$ accounts for at least some
of the tilt of the FP in the {\it r} band relative to the virial
theorem expectation. An increasing dark matter fraction with
increasing stellar mass could be responsible for this dynamical
mass-to-light ratio variation \citep{BMQ05}.

\begin{figure}
\includegraphics{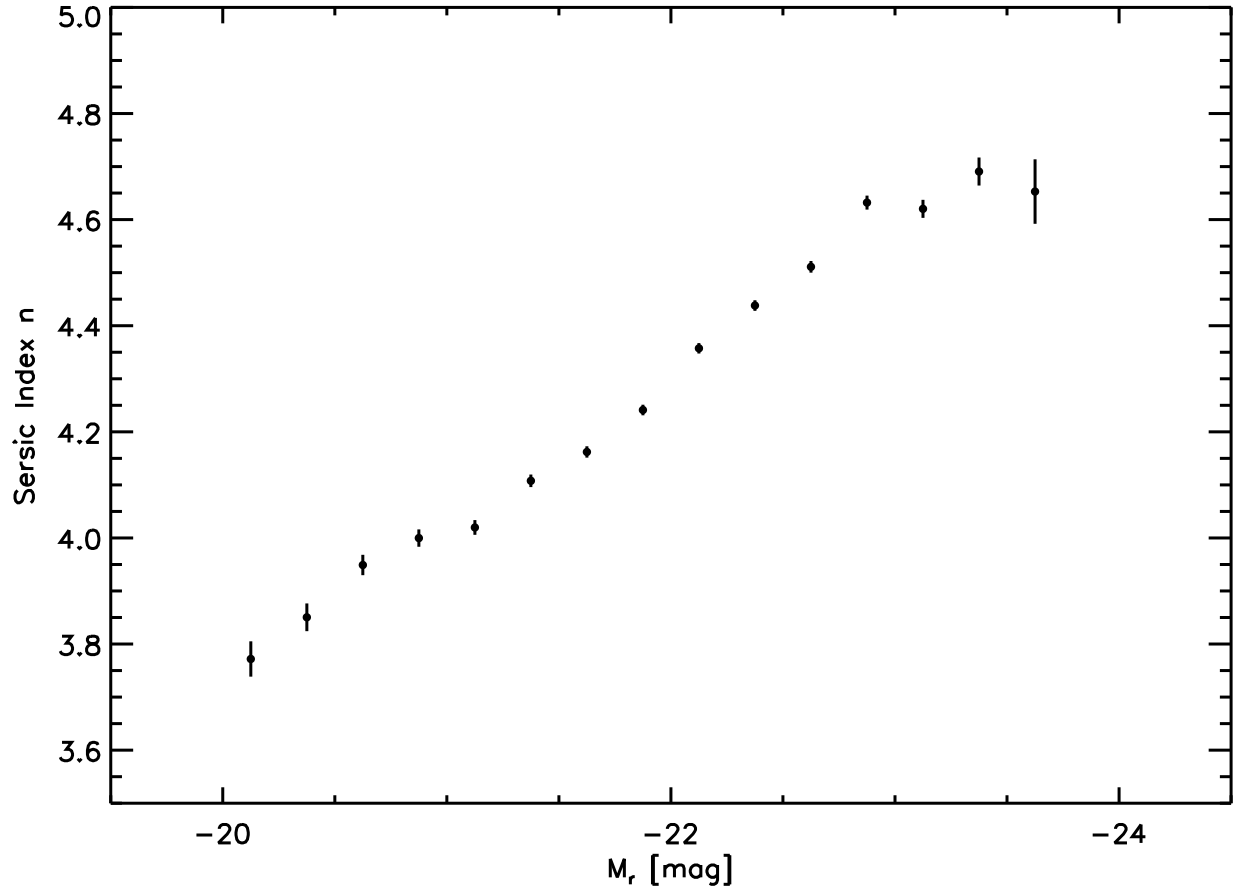}
\caption{Distributions of S\'ersic index in 0.25 mag wide bins for our non-BCG elliptical galaxy sample. Points
denote the mean of a gaussian fit to the distribution in a given bin,
while the error bars correspond to the $1\sigma$ error on the mean. 
The mild variation in index with luminosity for
low-luminosity ellipticals implies a non-homology within the
elliptical galaxy population.}
\label{fig:sersicn}
\end{figure}

As a check on the importance of non-homology for tilting the FP,
Figure~\ref{fig:sersicn} shows the S\'ersic index $n$ as a function of
luminosity for our sample.  Although the range of $n$ at any given
luminosity is quite large, there is a clear trend of increasing $n$
for more luminous galaxies, as has been found by other authors
\citep{Caon93,Graham96,Ferrarese06}.  This trend is slightly
underestimated in the VAGC due to the systematic problems associated
with the S\'ersic fits, which result in fluxes, radii, and S\'ersic
indices being underestimated by $\sim$10\% at $n=5$ \citep{Blanton05}.  
 The variation in $n$ we find is significantly less than
what \citet{Ciotti96} find
is required for non-homology to fully account for the tilt in the FP
across the entire luminosity range of early type galaxies. It is therefore likely that non-homology, while present, does not fully account for the tilt in the FP.

Aside from $L$ vs. $M_{\rm dyn}$, all of the other FP projections
exhibit a local slope that varies systematically with luminosity from
the faint end of the sample to the bright end.  The slope $\alpha$ of the
radius-luminosity relation for elliptical galaxies varies
systematically with luminosity, from $\alpha \approx 0.5$ at $M_r
\approx -20$ to $\alpha \approx 0.7$ at $M_r \approx -24$; for the
entire sample, we find $\alpha \approx 0.6$, in good agreement with
\citet{Shen03} and \citet{Bernardi03b}.  Our slope at faint luminosity of
$\alpha \approx 0.5$ agrees with that found by \citet{Lauer06a} for
their lower-luminosity ellipticals.  At the luminous end, however,
\citet{Lauer06a} find $\alpha \approx 1.1$, much steeper than we see,
though their sample consists primarily of BCGs in the {\it V} band. \citet{Bernardi06} attribute curvature in the radius-luminosity relationship at high luminosity in the full SDSS sample to an increasing fraction of BCGs dominating the sample. Our results show, however, that the normal non-BCG population exhibits a steepening at high luminosity that does not appear to be attributable to contamination by BCGs (see \S \ref{sec:data_bcg}).

The slope $\beta$ of the FJ relation also steepens from $\beta \approx 3$ at
$M_r \approx -20$ to $\beta \approx 4.5$ at $M_r \approx -24$; the
canonical FJ slope of $L\propto\sigma^4$ is found in the middle of our
sample, at $M_r=-22.5$.  Our slope at faint luminosity for the FJ relation is
similar to that found by \citet{Tonry81}, but not as flat as that
found by \citet{MG05}, who report $L\propto\sigma^{2.01 \pm 0.36}$ for
a sample of faint early-type galaxies; their sample, however, extends
to much fainter magnitudes ($-22.0 < M_r <-17.5$ mag).  At the
luminous end, our value of $\beta$ is significantly smaller than the
value of $\beta \approx 8$ found by \citealt{Lauer06a}, although this
is again a result of their sample being primarily BCGs.

The density-luminosity and surface brightness-luminosity relations
shown in Figures~\ref{fig:dists}-~\ref{fig:quad} indicate that
low-luminosity ellipticals have roughly constant densities and surface
brightnesses, while more luminous ellipticals become progressively
less dense and have lower surface brightness with increasing
luminosity.

As discussed in \S \ref{sec:data}, we find similar results for the FP
projections regardless of the magnitude we use to define the sample
(Petrosian, S\'ersic, or de Vaucouleurs). More specifically, the local
slopes of the quadratic fit to the FJ relation are nearly identical
for all three magnitude types.  For the radius-luminosity relation, de
Vaucouleurs fits result in $\alpha \approx 0.45$ at the faint end and
$\alpha \approx 0.75$ at the bright end, whereas with S\'ersic fits the
slope ranges from $\alpha \approx 0.5$ to $\alpha \approx 0.8$, both
similar to our default analysis (Fig. \ref{fig:quad}).

\subsection{Brightest Cluster Galaxies}
\label{sec:disc_bcg}

\begin{figure}
\includegraphics{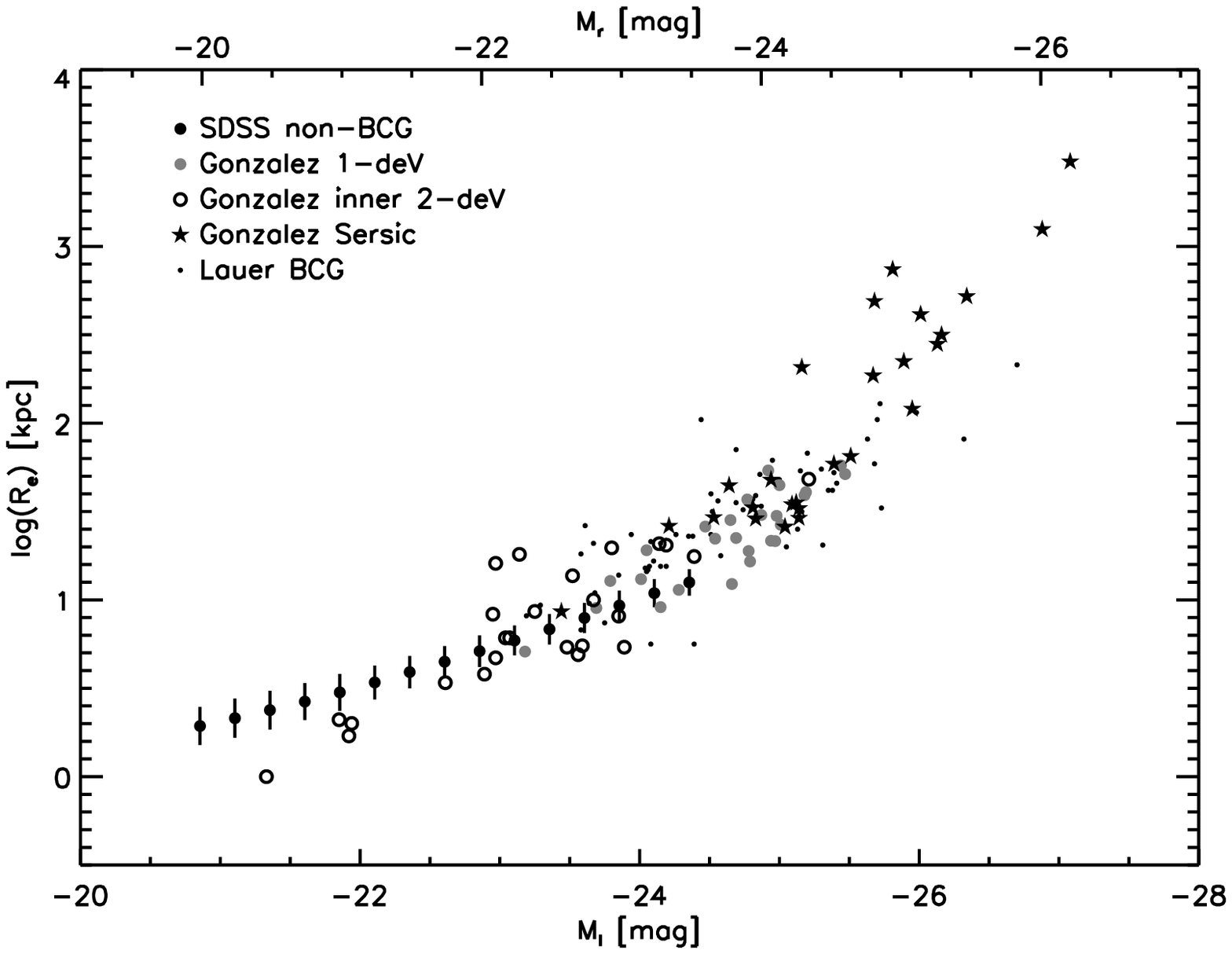}
\caption{Comparison between our non-BCG SDSS sample and the local BCG samples
of \citet{Gonzalez05} and \citet{Lauer06b} in the $I_c$ band. The corresponding {\it r} band magnitude is estimated using the median colour of {\it r-}$I_c=0.87$ for the SDSS sample. SDSS data (black circles) are
the effective radii vs. luminosity distributions in the {\it i} band, with BCGs excluded,
shifted to the $I_c$ band of Gonzalez using a median {\it i-z} colour
of 0.22 and the transformation equations obtained from the SDSS
website.  Also shown are Gonzalez's one component de Vaucouleurs fits
for the effective radius and luminosity of local BCGs (grey circles),
the inner component of Gonzalez's two component de Vaucouleurs fits to
the photometry of BCGs (open circles), and the radii and luminosities
from S\'ersic fits (stars).  The \citet{Lauer06b} BCGs (dots) are one component de Vaucouleurs fits, and have been shifted to the $I_c$ band using a median colour of $V$-$I=1.4$. The two de Vaucouleurs samples from
\citet{Gonzalez05} can be fit by $R_e \propto L^{1.0 \pm 0.1}$, which
is steeper than the $R_e \propto L^{0.8}$ correlation of normal
elliptical galaxies at the bright end in {\it i} band. The
S\'ersic sample is even steeper, with $R_e \propto L^{1.8 \pm
0.2}$.}
\label{fig:gonzalez}
\end{figure}

BCGs are at the extreme end of the luminous elliptical galaxy
population; it is thus interesting to compare our results on the FP
projections for luminous ellipticals to those of BCGs.  Figure
\ref{fig:gonzalez} shows our non-BCG SDSS data for the $R_e$-$L$ correlation
of ellipticals, along with a local sample of BCGs studied by
\citet{Gonzalez05}, who extend previous BCG studies by studying the
very extended low surface brightness emission at large radii and
comparing a variety of full two dimensional models for BCG surface
brightness profiles. We also include the local BCG sample of \citet{Lauer06b}.
We use SDSS {\it i} band data shifted to the $I_c$
band of \citeauthor{Gonzalez05} using the median {\it i-z} colour and
the transformation equations obtained from the SDSS website. (For ease of comparison, we also include the approximate {\it r} band magnitudes on the upper axis, using the median colour of {\it r-}$I_c=0.87$ for the SDSS sample.) The \citet{Lauer06b} points are shown as dots, shifted to the $I_c$ band using a median colour of $V$-$I=1.4$ for this sample. The solid
grey points in Figure \ref{fig:gonzalez} are \citeauthor{Gonzalez05}'s
1-component de Vaucouleurs fits for $R_e$ and $L$.  They also argue,
however, that S\'ersic fits and fits with two de Vaucouleurs
components (an inner and outer component) provide a much better
description of the surface brightness profiles of BCGs; the resulting
$R_e$ and $L$ are shown by stars and open circles in Figure
\ref{fig:gonzalez}, respectively.  For the two-component de
Vaucouleurs fits, \citet{Gonzalez05} argue that the inner component
may correspond to a ``normal'' elliptical galaxy while the outer
component represents an extended envelope.  Figure \ref{fig:gonzalez}
shows, however, that most of the effective radii for Gonzalez's inner
component fits lie well outside the observed distribution of $R_e$ for
normal ellipticals. This, together with the large scatter in the
$R_e$-$L$ relation for these inner component fits, makes it somewhat
difficult to physically interpret the two component model for BCG
surface brightness profiles.

\begin{figure}
\includegraphics{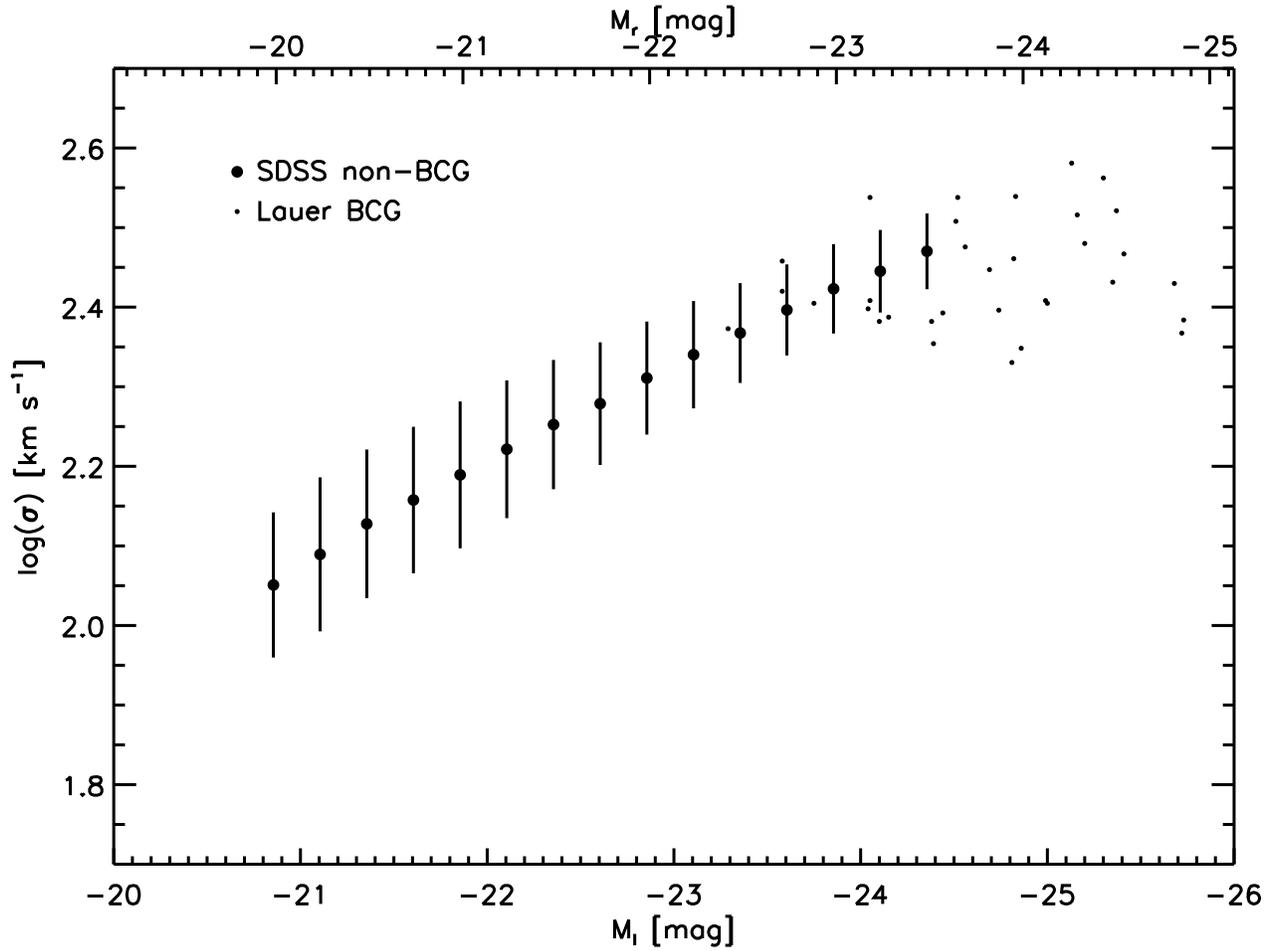}
\caption{Comparison between our non-BCG SDSS sample and the local BCG sample
of \citet{Lauer06a} in the $I_c$ band. The corresponding {\it r} band magnitude is estimated using the median colour of {\it r-}$I_c=0.87$ for the SDSS sample. SDSS data (black circles) are
the velocity dispersion vs. luminosity distributions in the {\it i} band, with BCGs excluded,
shifted to the $I_c$ band using a median {\it i-z} colour
of 0.22 and the transformation equations obtained from the SDSS
website.  The BCG data points from \citet{Lauer06a} have been shifted to the $I_c$ band using a median colour of $V$-$I=1.4$.}
\label{fig:Lsigma_lauer}
\end{figure}

The 1-component de Vaucouleurs fits for $R_e$ and $L$ from
\citet{Gonzalez05} yield an $R_e$-$L$ correlation of $R_e \propto L^{1
\pm 0.1}$.  \citet{Lauer06a} find a similar result for BCGs, as do
\citet{Bernardi06}. For comparison, we find that $R_e \propto L^{0.8}$
at the luminous end of the normal elliptical galaxy population in the
{\it i} band (note that this is slightly steeper than our {\it r} band
slope). In Figure~\ref{fig:Lsigma_lauer} we show the Faber-Jackson relation with our SDSS sample and the local BCG sample of \citet{Lauer06a}, both shifted to the $I_c$ band as described above. \citet{Lauer06a} find $L
\propto \sigma^8$ for local BCGs, while we find $L \propto
\sigma^{4.5}$ for the most luminous ellipticals.  The Faber-Jackson
and $R_e$-$L$ scalings for BCGs are thus significantly steeper than
those for normal elliptical galaxies, even at the luminous end of the
latter population. This is consistent with significant growth of BCGs
via dissipationless mergers (see \S \ref{sec:disc}, also \citealt{BMQ06}).

\begin{figure}
\includegraphics{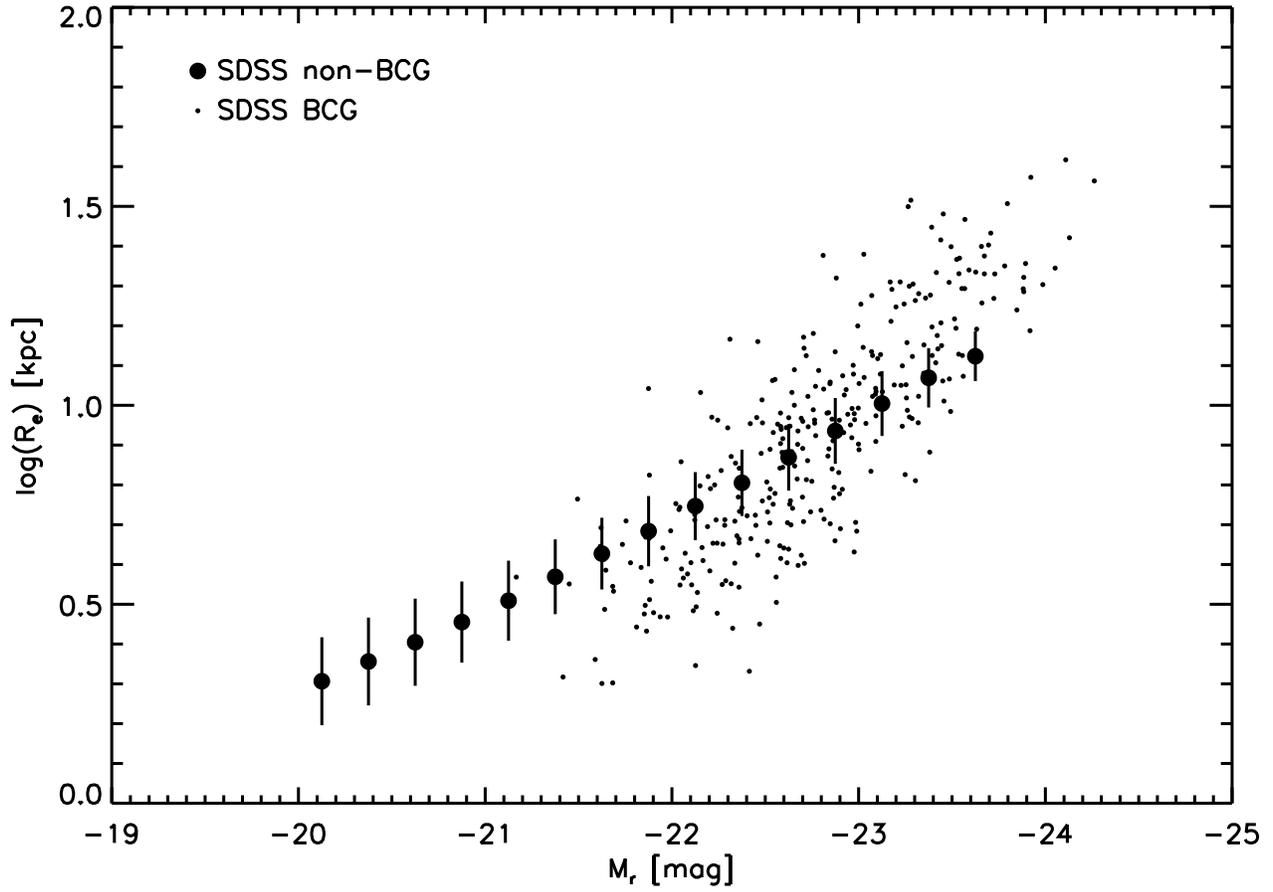}
\caption{Comparison of $R_e$ vs. $L$ for our non-BCG SDSS sample and the C4 BCGs, whose photometry has been reanalyzed using the sky subtraction method of \citet{West06}. The BCGs are disjoint from the non-BCGs, with a larger $R_e$-$L$ slope, and do not smoothly transition into the non-BCG relation.}
\label{fig:RL_allsdss}
\end{figure}

\begin{figure}
\includegraphics{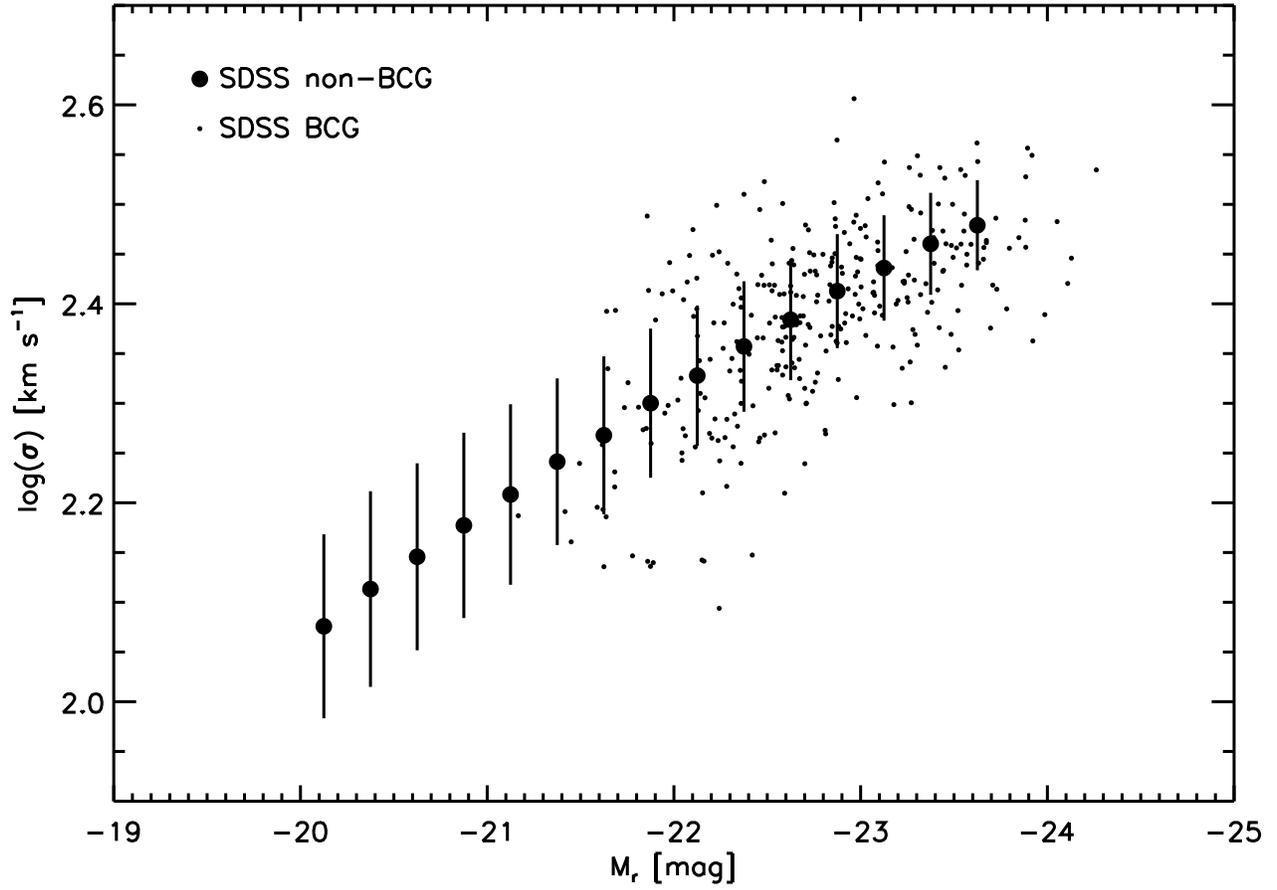}
\caption{Comparison of $\sigma$ vs. $L$ for our non-BCG SDSS sample and the C4 BCGs, whose photometry has been reanalyzed using the sky subtraction method of \citet{West06}. The BCGs do not clearly differentiate themselves from the non-BCGs, although they exhibit strong variations in slope at the high luminosity end.}
\label{fig:SL_allsdss}
\end{figure}

In Figures~\ref{fig:RL_allsdss} \&~\ref{fig:SL_allsdss} we compare the {\it r}-band radius-luminosity and FJ relations of our non-BCG main sample to our reanalysis of the C4 BCGs. The BCGs were reanalyzed as described in \S \ref{sec:data_bcg}, using the sky subtraction method of \citet{West06}. The radius-luminosity relation for BCGs appears distinct from the non-BCGs, exhibiting a steeper slope of $\alpha \approx 1.1$. By contrast, the FJ relation for BCGs appears to smoothly match the non-BCG sample, although it flattens significantly above $M_r=-23$. The overall FJ slope for the BCGs is $\beta \approx 4.2$, similar to the slope at the luminous end of the non-BCG sample. Brighter than $M_r=-23$, however, the FJ slope for BCGs is $\beta \approx 5.9$, while fainter than $M_r=-23$, the slope is $\beta \approx 3.0$. These results agree reasonably well with \citet{Bernardi06} and \citet{Lauer06a}, but \citet{vonderLinden06} find a much shallower radius-luminosity relation of $R_e \propto L^{0.65 \pm 0.02}$ for BCGs, consistent with normal elliptical galaxies. The latter sample is based on a careful reanalysis of C4 clusters, but with definitions of radius and luminosity that differ from those used here. 

Most of the luminosity in BCGs arises in extended low surface
brightness emission (e.g., \citealt{Gonzalez05}).  Thus observations of
different depths will yield different luminosities and the inferred
``galaxy'' luminosity and radius hinge critically on how much of the
extended low surface brightness emission is attributed to
intra-cluster light rather than galaxy light.  The importance of this
is illustrated by \citeauthor{Gonzalez05}'s S\'ersic fits in Figure
\ref{fig:gonzalez}, which give an $R_e$-$L$ correlation of $R_e \propto
L^{1.8 \pm 0.2}$, very different from the $R_e \propto L^{1 \pm 0.1}$
correlation of their one component de Vaucouleurs fits.  This
highlights the subtlety in defining the radii and luminosity of BCGs,
which needs to be taken into account when comparing the properties of
normal ellipticals and BCGs.

\section{DISCUSSION}
\label{sec:disc}

We have analyzed the fundamental plane projections of elliptical
galaxies as a function of luminosity, using a sample of $\approx
80,000$ galaxies drawn from DR4 of the SDSS.  We have separated BCGs
from our main elliptical galaxy sample and reanalyzed their photometry.
The observables we consider are
effective radius ($R_e$), velocity dispersion ($\sigma$), dynamical
mass ($M_{\rm dyn} \propto R_e\sigma^2$), effective density
($\sigma^2/R_{\!e}^2$), and effective surface brightness ($\mu_e$).
With the exception of the $L$-$M_{\rm dyn}$ correlation, we find clear
evidence of variations in the slope of the FP projections with
luminosity (Figs. \ref{fig:allslopes} \& \ref{fig:quad}) in the normal elliptical galaxy sample.
The trends
we find are that the radius-luminosity and Faber-Jackson relations are
steeper at high luminosity relative to low luminosity and that more
luminous ellipticals become progressively less dense and have lower
surface brightnesses than lower luminosity ellipticals.  These trends
are consistent with existing results in the literature, though our
results have much better statistics.  It is interesting to note that
the results we find appear to continue to even lower luminosity
spheroidal systems. In particular, for a sample of dwarf elliptical
(dE) and dwarf spheroidal (dSph) galaxies in the {\it B} band, the
radius-luminosity power-law slope $\alpha \approx$ 0.28--0.55 and the
Faber-Jackson power-law slope $\beta \approx$ 1.5--2.5
\citep{deRijcke05}, broadly consistent with an extrapolation of our
results to even lower luminosities. These trends are also seen in the dE sample of \citet{GG03}. We note, however, that there are likely structural differences between luminous ellipticals and dEs that complicate this simple extrapolation.

The variations in the FP projections we find at the
luminous end are consistent with less and less dissipation during the
formation of elliptical galaxies with increasing luminosity.  Less
dissipation would result in increasingly larger, less dense, lower
surface brightness, and lower $\sigma$ galaxies, relative to fiducial
power-law scalings for the elliptical galaxy population.  This
interpretation is consistent with a variety of other observational
evidence for less dissipation in the formation of luminous ellipticals
(e.g., \citealt{KB96}).

Luminosity (or galaxy-mass) dependent variations in the FP projections
should in principle provide a strong constraint on galaxy formation
models.  For example, numerical simulations of gas-free (``dry'')
merger remnants find that the FP projections have $\alpha \sim 0.7$
and $\beta \sim 4$ for very wide orbits, while lower angular momentum
and higher energy orbits have less energy exchange between stars and
dark matter and thus the FP projections steepen to $\alpha \sim 1$ and
$\beta \sim 6$ \citep{BMQ06}.  These values of $\alpha$ and $\beta$
are consistent with the trends we find in the data at the luminous
end, suggesting an increasing importance of dry mergers for luminous
ellipticals.  For lower-luminosity ellipticals, however, dry mergers
are less likely to be important in determining the properties of
elliptical galaxies.  Current simulations of elliptical galaxy
formation via mergers of gas-rich disks find that the remnants lie
roughly on the FP and its projections, with the gas fraction of the
progenitor galaxies strongly affecting the properties of the resulting
spheroidal merger remnants \citep{Robertson06}.  These simulations
also find more dissipation in the formation of lower luminosity
ellipticals, which is consistent with our data.  It remains to be seen
whether simulations of gas-rich mergers can also account for the
systematic variations in the FP projections with luminosity (e.g.,
with a varying gas fraction).

We have also reanalyzed the photometry of SDSS BCGs using the C4 cluster sample of \citet{Miller05}; the standard pipeline photometry underestimates the luminosities and radii for luminous BCGs (Figure~\ref{fig:RevsRe}) due to an overestimate of sky background.  With our corrected photometry, we find that the radius-luminosity relation of BCGs is noticeably steeper than that of normal (non-BCG) ellipticals and the FJ relation for BCGs flattens at the luminous end. The same trends are also seen in local samples of BCGs (eg., \citealt{Lauer06a}).  These observational results are consistent with significant growth of BCGs by dissipationless mergers. 

\subsection{Massive Black Holes}

For the bulk of the elliptical galaxy population, both the $M_{\rm
BH}$-$\sigma$ and Faber-Jackson relations scale roughly as $\sigma^4$.
As noted in \S 1, however, \citet{Wyithe06} argued that the $M_{\rm
BH}$-$\sigma$ relation steepens at high black hole masses and flattens
at low masses, albeit with only modest statistical significance.  \citet{GH06} see a similar flattening at low black
hole masses and \citet{Lauer06a} argue for a similar steepening at high
black hole masses based on the properties of the surface brightness
cores in luminous elliptical galaxies.  These luminosity-dependent
variations in the $M_{\rm BH}$-$\sigma$ relation are qualitatively
similar to the luminosity dependent variation in the FJ relation.
Together with a constant $M_\star/L$, these results imply a roughly
constant value of $M_{\rm BH}/M_\star$ across a range of early type
galaxies (see, e.g., \citealt{HR04} for a direct compilation of
$M_{\rm BH}/M_\star$ that bears this out).  For massive galaxies, the
approximate constancy of $M_{\rm BH}/M_\star$ may be a consequence of
dissipationless mergers, which preserve $M_{\rm BH}/M_\star$ in the
absence of energy loss by gravitational waves during black hole
coalescence \citep{BMQ06}.  For lower luminosity ellipticals, however,
gas dynamics is important and a constant value of $M_{\rm BH}/M_\star$
implies a constant relative efficiency for forming stars and massive
black holes in a given gravitational potential well.

\section*{ACKNOWLEDGMENTS}

We thank Michael Blanton and David Schlegel for their advice and help
with understanding the details of the VAGC. We also thank Tod Lauer,
Michael Boylan-Kolchin, Alister Graham, Robert Lupton, David Fisher, and an anonymous referee for helpful
conversations and comments. L.-B.D. is grateful for a Julie-Payette/Doctoral
Fellowship from the Natural Sciences and Engineering Research Council
of Canada and by the Canadian Space Agency.  E.Q. is supported in part
by NASA grant NNG05GO22H, the Alfred P. Sloan Foundation and the David
and Lucile Packard Foundation. C.-P.M. is supported in part by NSF
grant AST-0407351 and NASA grant NAG5-12173. A.A.W. is supported by
NSF grant AST-0540567

Funding for the SDSS and SDSS-II has been provided by the Alfred P. Sloan Foundation, the Participating Institutions, the National Science Foundation, the U.S. Department of Energy, the National Aeronautics and Space Administration, the Japanese Monbukagakusho, the Max Planck Society, and the Higher Education Funding Council for England. The SDSS Web Site is http://www.sdss.org/.

The SDSS is managed by the Astrophysical Research Consortium for the Participating Institutions. The Participating Institutions are the American Museum of Natural History, Astrophysical Institute Potsdam, University of Basel, University of Cambridge, Case Western Reserve University, University of Chicago, Drexel University, Fermilab, the Institute for Advanced Study, the Japan Participation Group, Johns Hopkins University, the Joint Institute for Nuclear Astrophysics, the Kavli Institute for Particle Astrophysics and Cosmology, the Korean Scientist Group, the Chinese Academy of Sciences (LAMOST), Los Alamos National Laboratory, the Max-Planck-Institute for Astronomy (MPIA), the Max-Planck-Institute for Astrophysics (MPA), New Mexico State University, Ohio State University, University of Pittsburgh, University of Portsmouth, Princeton University, the United States Naval Observatory, and the University of Washington.

\bibliographystyle{apj}
\bibliography{mybiblio}

\begin{thebibliography}{65}
\expandafter\ifx\csname natexlab\endcsname\relax\def\natexlab#1{#1}\fi

\bibitem[{{Adelman-McCarthy} {et~al.}(2006){Adelman-McCarthy}, {Ag{\"u}eros},
  {Allam}, {Anderson}, {Anderson}, {Annis}, {Bahcall}, {Baldry}, {Barentine},
  {Berlind}, {Bernardi}, {Blanton}, {Boroski}, {Brewington}, {Brinchmann},
  {Brinkmann}, {Brunner}, {Budav{\'a}ri}, {Carey}, {Carr}, {Castander},
  {Connolly}, {Csabai}, {Czarapata}, {Dalcanton}, {Doi}, {Dong}, {Eisenstein},
  {Evans}, {Fan}, {Finkbeiner}, {Friedman}, {Frieman}, {Fukugita}, {Gillespie},
  {Glazebrook}, {Gray}, {Grebel}, {Gunn}, {Gurbani}, {de Haas}, {Hall},
  {Harris}, {Harvanek}, {Hawley}, {Hayes}, {Hendry}, {Hennessy}, {Hindsley},
  {Hirata}, {Hogan}, {Hogg}, {Holmgren}, {Holtzman}, {Ichikawa}, {Ivezi{\'c}},
  {Jester}, {Johnston}, {Jorgensen}, {Juri{\'c}}, {Kent}, {Kleinman}, {Knapp},
  {Kniazev}, {Kron}, {Krzesinski}, {Kuropatkin}, {Lamb}, {Lampeitl}, {Lee},
  {Leger}, {Lin}, {Long}, {Loveday}, {Lupton}, {Margon},
  {Mart{\'{\i}}nez-Delgado}, {Mandelbaum}, {Matsubara}, {McGehee}, {McKay},
  {Meiksin}, {Munn}, {Nakajima}, {Nash}, {Neilsen}, {Newberg}, {Newman},
  {Nichol}, {Nicinski}, {Nieto-Santisteban}, {Nitta}, {O'Mullane}, {Okamura},
  {Owen}, {Padmanabhan}, {Pauls}, {Peoples}, {Pier}, {Pope}, {Pourbaix},
  {Quinn}, {Richards}, {Richmond}, {Rockosi}, {Schlegel}, {Schneider},
  {Schroeder}, {Scranton}, {Seljak}, {Sheldon}, {Shimasaku}, {Smith}, {Smol{\v
  c}i{\'c}}, {Snedden}, {Stoughton}, {Strauss}, {SubbaRao}, {Szalay},
  {Szapudi}, {Szkody}, {Tegmark}, {Thakar}, {Tucker}, {Uomoto}, {Vanden Berk},
  {Vandenberg}, {Vogeley}, {Voges}, {Vogt}, {Walkowicz}, {Weinberg}, {West},
  {White}, {Xu}, {Yanny}, {Yocum}, {York}, {Zehavi}, {Zibetti}, \&
  {Zucker}}]{sdss_dr4}
{Adelman-McCarthy}, J.~K. {et~al.} 2006, ApJS, 162, 38

\bibitem[{{Bender} {et~al.}(1992){Bender}, {Burstein}, \& {Faber}}]{BBF92}
{Bender}, R., {Burstein}, D., \& {Faber}, S.~M. 1992, ApJ, 399, 462

\bibitem[{{Bernardi} {et~al.}(2006){Bernardi}, {Hyde}, {Sheth}, {Miller}, \&
  {Nichol}}]{Bernardi06}
{Bernardi}, M., {Hyde}, J.~B., {Sheth}, R.~K., {Miller}, C.~J., \& {Nichol},
  R.~C. 2006, AJ, in press (astro-ph/0607117)

\bibitem[{{Bernardi} {et~al.}(2003{\natexlab{a}}){Bernardi}, {Sheth}, {Annis},
  {Burles}, {Eisenstein}, {Finkbeiner}, {Hogg}, {Lupton}, {Schlegel},
  {SubbaRao}, {Bahcall}, {Blakeslee}, {Brinkmann}, {Castander}, {Connolly},
  {Csabai}, {Doi}, {Fukugita}, {Frieman}, {Heckman}, {Hennessy}, {Ivezi{\'c}},
  {Knapp}, {Lamb}, {McKay}, {Munn}, {Nichol}, {Okamura}, {Schneider}, {Thakar},
  \& {York}}]{Bernardi03c}
{Bernardi}, M. {et~al.} 2003{\natexlab{a}}, AJ, 125, 1866

\bibitem[{{Bernardi} {et~al.}(2003{\natexlab{b}}){Bernardi}, {Sheth}, {Annis},
  {Burles}, {Eisenstein}, {Finkbeiner}, {Hogg}, {Lupton}, {Schlegel},
  {SubbaRao}, {Bahcall}, {Blakeslee}, {Brinkmann}, {Castander}, {Connolly},
  {Csabai}, {Doi}, {Fukugita}, {Frieman}, {Heckman}, {Hennessy}, {Ivezi{\'c}},
  {Knapp}, {Lamb}, {McKay}, {Munn}, {Nichol}, {Okamura}, {Schneider}, {Thakar},
  \& {York}}]{Bernardi03b}
---. 2003{\natexlab{b}}, AJ, 125, 1849

\bibitem[{{Bernardi} {et~al.}(2003{\natexlab{c}}){Bernardi}, {Sheth}, {Annis},
  {Burles}, {Eisenstein}, {Finkbeiner}, {Hogg}, {Lupton}, {Schlegel},
  {SubbaRao}, {Bahcall}, {Blakeslee}, {Brinkmann}, {Castander}, {Connolly},
  {Csabai}, {Doi}, {Fukugita}, {Frieman}, {Heckman}, {Hennessy}, {Ivezi{\'c}},
  {Knapp}, {Lamb}, {McKay}, {Munn}, {Nichol}, {Okamura}, {Schneider}, {Thakar},
  \& {York}}]{Bernardi03a}
---. 2003{\natexlab{c}}, AJ, 125, 1817

\bibitem[{{Blanton} {et~al.}(2003{\natexlab{a}}){Blanton}, {Brinkmann},
  {Csabai}, {Doi}, {Eisenstein}, {Fukugita}, {Gunn}, {Hogg}, \&
  {Schlegel}}]{Blanton03_aj125}
{Blanton}, M.~R. {et~al.} 2003{\natexlab{a}}, AJ, 125, 2348

\bibitem[{{Blanton} {et~al.}(2003{\natexlab{b}}){Blanton}, {Hogg}, {Bahcall},
  {Baldry}, {Brinkmann}, {Csabai}, {Eisenstein}, {Fukugita}, {Gunn},
  {Ivezi{\'c}}, {Lamb}, {Lupton}, {Loveday}, {Munn}, {Nichol}, {Okamura},
  {Schlegel}, {Shimasaku}, {Strauss}, {Vogeley}, \&
  {Weinberg}}]{Blanton03_apj594}
---. 2003{\natexlab{b}}, ApJ, 594, 186

\bibitem[{{Blanton} {et~al.}(2003{\natexlab{c}}){Blanton}, {Hogg}, {Bahcall},
  {Brinkmann}, {Britton}, {Connolly}, {Csabai}, {Fukugita}, {Loveday},
  {Meiksin}, {Munn}, {Nichol}, {Okamura}, {Quinn}, {Schneider}, {Shimasaku},
  {Strauss}, {Tegmark}, {Vogeley}, \& {Weinberg}}]{Blanton03_apj592}
---. 2003{\natexlab{c}}, ApJ, 592, 819

\bibitem[{{Blanton} {et~al.}(2005){Blanton}, {Schlegel}, {Strauss},
  {Brinkmann}, {Finkbeiner}, {Fukugita}, {Gunn}, {Hogg}, {Ivezi{\'c}}, {Knapp},
  {Lupton}, {Munn}, {Schneider}, {Tegmark}, \& {Zehavi}}]{Blanton05}
---. 2005, AJ, 129, 2562

\bibitem[{{Boylan-Kolchin} {et~al.}(2005){Boylan-Kolchin}, {Ma}, \&
  {Quataert}}]{BMQ05}
{Boylan-Kolchin}, M., {Ma}, C.-P., \& {Quataert}, E. 2005, MNRAS, 362, 184

\bibitem[{{Boylan-Kolchin} {et~al.}(2006){Boylan-Kolchin}, {Ma}, \&
  {Quataert}}]{BMQ06}
---. 2006, MNRAS, 369, 1081

\bibitem[{{Burstein} {et~al.}(1997){Burstein}, {Bender}, {Faber}, \&
  {Nolthenius}}]{Burstein97}
{Burstein}, D., {Bender}, R., {Faber}, S., \& {Nolthenius}, R. 1997, AJ, 114,
  1365

\bibitem[{{Caon} {et~al.}(1993){Caon}, {Capaccioli}, \& {D'Onofrio}}]{Caon93}
{Caon}, N., {Capaccioli}, M., \& {D'Onofrio}, M. 1993, MNRAS, 265, 1013

\bibitem[{{Ciotti} {et~al.}(1996){Ciotti}, {Lanzoni}, \& {Renzini}}]{Ciotti96}
{Ciotti}, L., {Lanzoni}, B., \& {Renzini}, A. 1996, MNRAS, 282, 1

\bibitem[{{Davies} {et~al.}(1983){Davies}, {Efstathiou}, {Fall}, {Illingworth},
  \& {Schechter}}]{Davies83}
{Davies}, R.~L., {Efstathiou}, G., {Fall}, S.~M., {Illingworth}, G., \&
  {Schechter}, P.~L. 1983, ApJ, 266, 41

\bibitem[{{de Rijcke} {et~al.}(2005){de Rijcke}, {Michielsen}, {Dejonghe},
  {Zeilinger}, \& {Hau}}]{deRijcke05}
{de Rijcke}, S., {Michielsen}, D., {Dejonghe}, H., {Zeilinger}, W.~W., \&
  {Hau}, G.~K.~T. 2005, A\&A, 438, 491

\bibitem[{{de Vaucouleurs}(1948)}]{deV48}
{de Vaucouleurs}, G. 1948, Annales d'Astrophysique, 11, 247

\bibitem[{{Djorgovski} \& {Davis}(1987)}]{DD87}
{Djorgovski}, S., \& {Davis}, M. 1987, ApJ, 313, 59

\bibitem[{{Dressler} {et~al.}(1987){Dressler}, {Lynden-Bell}, {Burstein},
  {Davies}, {Faber}, {Terlevich}, \& {Wegner}}]{Dressler87}
{Dressler}, A., {Lynden-Bell}, D., {Burstein}, D., {Davies}, R.~L., {Faber},
  S.~M., {Terlevich}, R., \& {Wegner}, G. 1987, ApJ, 313, 42

\bibitem[{{Faber} \& {Jackson}(1976)}]{FJ76}
{Faber}, S.~M., \& {Jackson}, R.~E. 1976, ApJ, 204, 668

\bibitem[{{Ferrarese} {et~al.}(2006){Ferrarese}, {C{\^o}t{\'e}}, {Jord{\'a}n},
  {Peng}, {Blakeslee}, {Piatek}, {Mei}, {Merritt}, {Milosavljevi{\'c}},
  {Tonry}, \& {West}}]{Ferrarese06}
{Ferrarese}, L. {et~al.} 2006, ApJS, 164, 334

\bibitem[{{Ferrarese} \& {Merritt}(2000)}]{FM00}
{Ferrarese}, L., \& {Merritt}, D. 2000, ApJL, 539, L9

\bibitem[{{Ferrarese} {et~al.}(2001){Ferrarese}, {Pogge}, {Peterson},
  {Merritt}, {Wandel}, \& {Joseph}}]{Ferrarese01}
{Ferrarese}, L., {Pogge}, R.~W., {Peterson}, B.~M., {Merritt}, D., {Wandel},
  A., \& {Joseph}, C.~L. 2001, ApJL, 555, L79

\bibitem[{{Gebhardt} {et~al.}(2000){Gebhardt}, {Bender}, {Bower}, {Dressler},
  {Faber}, {Filippenko}, {Green}, {Grillmair}, {Ho}, {Kormendy}, {Lauer},
  {Magorrian}, {Pinkney}, {Richstone}, \& {Tremaine}}]{Gebhardt00}
{Gebhardt}, K. {et~al.} 2000, ApJL, 539, L13

\bibitem[{{Gerhard} {et~al.}(2001){Gerhard}, {Kronawitter}, {Saglia}, \&
  {Bender}}]{Gerhard01}
{Gerhard}, O., {Kronawitter}, A., {Saglia}, R.~P., \& {Bender}, R. 2001, AJ,
  121, 1936

\bibitem[{{Gonzalez} {et~al.}(2005){Gonzalez}, {Zabludoff}, \&
  {Zaritsky}}]{Gonzalez05}
{Gonzalez}, A.~H., {Zabludoff}, A.~I., \& {Zaritsky}, D. 2005, ApJ, 618, 195

\bibitem[{{Graham} \& {Colless}(1997)}]{GC97}
{Graham}, A., \& {Colless}, M. 1997, MNRAS, 287, 221

\bibitem[{{Graham} {et~al.}(1996){Graham}, {Lauer}, {Colless}, \&
  {Postman}}]{Graham96}
{Graham}, A., {Lauer}, T.~R., {Colless}, M., \& {Postman}, M. 1996, ApJ, 465,
  534

\bibitem[{{Graham} {et~al.}(2005){Graham}, {Driver}, {Petrosian}, {Conselice},
  {Bershady}, {Crawford}, \& {Goto}}]{Graham05}
{Graham}, A.~W., {Driver}, S.~P., {Petrosian}, V., {Conselice}, C.~J.,
  {Bershady}, M.~A., {Crawford}, S.~M., \& {Goto}, T. 2005, AJ, 130, 1535

\bibitem[{{Graham} {et~al.}(2001){Graham}, {Erwin}, {Caon}, \&
  {Trujillo}}]{Graham01}
{Graham}, A.~W., {Erwin}, P., {Caon}, N., \& {Trujillo}, I. 2001, ApJL, 563,
  L11

\bibitem[{{Graham} \& {Guzm{\'a}n}(2003)}]{GG03}
{Graham}, A.~W., \& {Guzm{\'a}n}, R. 2003, AJ, 125, 2936

\bibitem[{{Greene} \& {Ho}(2006)}]{GH06}
{Greene}, J.~E., \& {Ho}, L.~C. 2006, ApJL, 641, L21

\bibitem[{{H{\"a}ring} \& {Rix}(2004)}]{HR04}
{H{\"a}ring}, N., \& {Rix}, H.-W. 2004, ApJL, 604, L89

\bibitem[{{J\o rgensen} {et~al.}(1995){J\o rgensen}, {Franx}, \&
  {Kjaergaard}}]{JFK95}
{J\o rgensen}, I., {Franx}, M., \& {Kjaergaard}, P. 1995, MNRAS, 276, 1341

\bibitem[{{Kauffmann} {et~al.}(2003){Kauffmann}, {Heckman}, {White}, {Charlot},
  {Tremonti}, {Brinchmann}, {Bruzual}, {Peng}, {Seibert}, {Bernardi},
  {Blanton}, {Brinkmann}, {Castander}, {Cs{\'a}bai}, {Fukugita}, {Ivezic},
  {Munn}, {Nichol}, {Padmanabhan}, {Thakar}, {Weinberg}, \&
  {York}}]{Kauffmann03}
{Kauffmann}, G. {et~al.} 2003, MNRAS, 341, 33

\bibitem[{{Kormendy} \& {Bender}(1996)}]{KB96}
{Kormendy}, J., \& {Bender}, R. 1996, ApJL, 464, L119+

\bibitem[{{Lauer} {et~al.}(2006{\natexlab{a}}){Lauer}, {Faber}, {Richstone},
  {Gebhardt}, {Tremaine}, {Postman}, {Dressler}, {Aller}, {Filippenko},
  {Green}, {Ho}, {Kormendy}, {Magorrian}, \& {Pinkney}}]{Lauer06a}
{Lauer}, T.~R. {et~al.} 2006{\natexlab{a}}, ApJ, in press (astro-ph/0606739)

\bibitem[{{Lauer} {et~al.}(2006{\natexlab{b}}){Lauer}, {Gebhardt}, {Faber},
  {Richstone}, {Tremaine}, {Kormendy}, {Aller}, {Bender}, {Dressler},
  {Filippenko}, {Green}, \& {Ho}}]{Lauer06b}
---. 2006{\natexlab{b}}, ApJ, in press (astro-ph/0609762)

\bibitem[{{Lin} {et~al.}(1999){Lin}, {Yee}, {Carlberg}, {Morris}, {Sawicki},
  {Patton}, {Wirth}, \& {Shepherd}}]{Lin99}
{Lin}, H., {Yee}, H.~K.~C., {Carlberg}, R.~G., {Morris}, S.~L., {Sawicki}, M.,
  {Patton}, D.~R., {Wirth}, G., \& {Shepherd}, C.~W. 1999, ApJ, 518, 533

\bibitem[{{Magorrian} {et~al.}(1998){Magorrian}, {Tremaine}, {Richstone},
  {Bender}, {Bower}, {Dressler}, {Faber}, {Gebhardt}, {Green}, {Grillmair},
  {Kormendy}, \& {Lauer}}]{Magorrian98}
{Magorrian}, J. {et~al.} 1998, AJ, 115, 2285

\bibitem[{{Marconi} \& {Hunt}(2003)}]{MH03}
{Marconi}, A., \& {Hunt}, L.~K. 2003, ApJL, 589, L21

\bibitem[{{Matkovi{\'c}} \& {Guzm{\'a}n}(2005)}]{MG05}
{Matkovi{\'c}}, A., \& {Guzm{\'a}n}, R. 2005, MNRAS, 362, 289

\bibitem[{{Miller} {et~al.}(2005){Miller}, {Nichol}, {Reichart}, {Wechsler},
  {Evrard}, {Annis}, {McKay}, {Bahcall}, {Bernardi}, {Boehringer}, {Connolly},
  {Goto}, {Kniazev}, {Lamb}, {Postman}, {Schneider}, {Sheth}, \&
  {Voges}}]{Miller05}
{Miller}, C.~J. {et~al.} 2005, AJ, 130, 968

\bibitem[{{Nakamura} {et~al.}(2003){Nakamura}, {Fukugita}, {Yasuda}, {Loveday},
  {Brinkmann}, {Schneider}, {Shimasaku}, \& {SubbaRao}}]{Nakamura03}
{Nakamura}, O., {Fukugita}, M., {Yasuda}, N., {Loveday}, J., {Brinkmann}, J.,
  {Schneider}, D.~P., {Shimasaku}, K., \& {SubbaRao}, M. 2003, AJ, 125, 1682

\bibitem[{{Oegerle} \& {Hoessel}(1991)}]{OH91}
{Oegerle}, W.~R., \& {Hoessel}, J.~G. 1991, ApJ, 375, 15

\bibitem[{{Oke} \& {Gunn}(1983)}]{OG83}
{Oke}, J.~B., \& {Gunn}, J.~E. 1983, ApJ, 266, 713

\bibitem[{{Padmanabhan} {et~al.}(2004){Padmanabhan}, {Seljak}, {Strauss},
  {Blanton}, {Kauffmann}, {Schlegel}, {Tremonti}, {Bahcall}, {Bernardi},
  {Brinkmann}, {Fukugita}, \& {Ivezi{\'c}}}]{Padmanabhan04}
{Padmanabhan}, N. {et~al.} 2004, New Astronomy, 9, 329

\bibitem[{{Petrosian}(1976)}]{Petrosian76}
{Petrosian}, V. 1976, ApJL, 209, L1

\bibitem[{{Postman} \& {Lauer}(1995)}]{PL95}
{Postman}, M., \& {Lauer}, T.~R. 1995, ApJ, 440, 28

\bibitem[{{Robertson} {et~al.}(2006){Robertson}, {Cox}, {Hernquist}, {Franx},
  {Hopkins}, {Martini}, \& {Springel}}]{Robertson06}
{Robertson}, B., {Cox}, T.~J., {Hernquist}, L., {Franx}, M., {Hopkins}, P.~F.,
  {Martini}, P., \& {Springel}, V. 2006, ApJ, 641, 21

\bibitem[{{S\'ersic}(1963)}]{Sersic63}
{S\'ersic}, J.~L. 1963, Boletin de la Asociacion Argentina de Astronomia La
  Plata Argentina, 6, 41

\bibitem[{{S\'ersic}(1968)}]{Sersic68}
---. 1968, {Atlas de galaxias australes} (Cordoba, Argentina: Observatorio
  Astronomico)

\bibitem[{{Shen} {et~al.}(2003){Shen}, {Mo}, {White}, {Blanton}, {Kauffmann},
  {Voges}, {Brinkmann}, \& {Csabai}}]{Shen03}
{Shen}, S., {Mo}, H.~J., {White}, S.~D.~M., {Blanton}, M.~R., {Kauffmann}, G.,
  {Voges}, W., {Brinkmann}, J., \& {Csabai}, I. 2003, MNRAS, 343, 978

\bibitem[{{Sheth} {et~al.}(2003){Sheth}, {Bernardi}, {Schechter}, {Burles},
  {Eisenstein}, {Finkbeiner}, {Frieman}, {Lupton}, {Schlegel}, {Subbarao},
  {Shimasaku}, {Bahcall}, {Brinkmann}, \& {Ivezi{\'c}}}]{Sheth03}
{Sheth}, R.~K. {et~al.} 2003, ApJ, 594, 225

\bibitem[{{Shimasaku} {et~al.}(2001){Shimasaku}, {Fukugita}, {Doi}, {Hamabe},
  {Ichikawa}, {Okamura}, {Sekiguchi}, {Yasuda}, {Brinkmann}, {Csabai},
  {Ichikawa}, {Ivezi{\'c}}, {Kunszt}, {Schneider}, {Szokoly}, {Watanabe}, \&
  {York}}]{Shimasaku01}
{Shimasaku}, K. {et~al.} 2001, AJ, 122, 1238

\bibitem[{{Tonry}(1981)}]{Tonry81}
{Tonry}, J.~L. 1981, ApJL, 251, L1

\bibitem[{{Tremaine} {et~al.}(2002){Tremaine}, {Gebhardt}, {Bender}, {Bower},
  {Dressler}, {Faber}, {Filippenko}, {Green}, {Grillmair}, {Ho}, {Kormendy},
  {Lauer}, {Magorrian}, {Pinkney}, \& {Richstone}}]{Tremaine02}
{Tremaine}, S. {et~al.} 2002, ApJ, 574, 740

\bibitem[{{Trujillo} {et~al.}(2004){Trujillo}, {Burkert}, \& {Bell}}]{TBB04}
{Trujillo}, I., {Burkert}, A., \& {Bell}, E.~F. 2004, ApJL, 600, L39

\bibitem[{{von der Linden} {et~al.}(2006){von der Linden}, {Best}, {Kauffmann},
  \& {White}}]{vonderLinden06}
{von der Linden}, A., {Best}, P.~N., {Kauffmann}, G., \& {White}, S.~D.~M.
  2006, MNRAS, in press (astro-ph/0611196)

\bibitem[{{Wake} {et~al.}(2006){Wake}, {Nichol}, {Eisenstein}, {Loveday},
  {Edge}, {Cannon}, {Smail}, {Schneider}, {Scranton}, {Carson}, {Ross},
  {Brunner}, {Colless}, {Couch}, {Croom}, {Driver}, {da Angela}, {Jester}, {de
  Propris}, {Drinkwater}, {Bland-Hawthorn}, {Pimbblet}, {Roseboom}, {Shanks},
  {Sharp}, \& {Brinkmann}}]{Wake06}
{Wake}, D.~A. {et~al.} 2006, MNRAS, in press (astro-ph/0607629)

\bibitem[{{Wegner} {et~al.}(1999){Wegner}, {Colless}, {Saglia}, {McMahan},
  {Davies}, {Burstein}, \& {Baggley}}]{Wegner99}
{Wegner}, G., {Colless}, M., {Saglia}, R.~P., {McMahan}, R.~K., {Davies},
  R.~L., {Burstein}, D., \& {Baggley}, G. 1999, MNRAS, 305, 259

\bibitem[{{West}(2005)}]{West05}
{West}, A.~A. 2005, Ph.D.~Thesis

\bibitem[{{West} {et~al.}(2007)}]{West06}
{West}, A.~A., {et~al.} 2007, AJ, submitted

\bibitem[{{Wyithe}(2006)}]{Wyithe06}
{Wyithe}, J.~S.~B. 2006, MNRAS, 365, 1082, {Erratum:} Wyithe, J.~S.~B. 2006,
  MNRAS, 371, 1536.

\end{thebibliography}

\begin{table}
\caption{Quadratic Fitting Parameters}
\label{tab:quad}
\begin{tabular}{lccc}
\hline
FP Projection & $a$ & $b$ & $c$ \\
\hline
$\log(R_e)$ vs. $\log(L)$ & 1.50 $\pm$ 0.19 & -0.802 $\pm$ 0.044 & 0.0805 $\pm$ 0.0026 \\
$\log(\sigma)$ vs. $\log(L)$ & -1.79 $\pm$ 0.23 & 0.674 $\pm$ 0.053 & -0.0234 $\pm$ 0.0030 \\
$\log(R_e\sigma^2)$ vs. $\log(L)$ & -3.37 $\pm$ 0.59 & 0.81 $\pm$ 0.14 & 0.0199 $\pm$ 0.0078 \\
$\log(\sigma/R_e)^2$ vs. $\log(L)$ & -4.82 $\pm$ 0.75 & 2.54 $\pm$ 0.17 & -0.1845 $\pm$ 0.0098 \\
$\mu_e$ vs. $\log(L)$ & 43.4 $\pm$ 2.2 & -5.77 $\pm$ 0.51 & 0.350 $\pm$ 0.030\\
\hline
\end{tabular}

\medskip
All fits are $X=a+b(\log(L))+c(\log(L))^2$, where $\log(L)$ is the
luminosity ($M=-2.5\log(L) $) and $X$ is the observable. $L$ is
calibrated to the AB system, in which a magnitude 0 object has the
same counts as a $F_\nu=3631$ Jy source; $\sigma$ is in units of km
s$^{-1}$; $R_e$ is in units of kpc; $\mu_e$ is in units of mag
arcsec$^{-2}$.
\end{table}

\clearpage

\clearpage

\clearpage

\clearpage

\clearpage

\clearpage

\clearpage

\clearpage

\clearpage

\end{document}